\documentclass{aa}  

\usepackage{graphicx}

\usepackage{txfonts}
\usepackage[final]{changes}

\begin{document} 

   \title{The Aperture Array Verification System 1: System overview and early commissioning results}

   \author{
P. Benthem \inst{1}
\and R. Wayth \inst{2}
\and E. de Lera Acedo \inst{12}
\and K. Zarb Adami \inst{11, 13}
\and M. Alderighi \inst{5}
\and C. Belli \inst{4}
\and P. Bolli  \inst{4}
\and T. Booler \inst{2}
\and J. Borg \inst{11}
\and J.~W.~Broderick \inst{2,1}
\and S. Chiarucci \inst{4}
\and R. Chiello \inst{13}
\and L. Ciani \inst{10}
\and G. Comoretto \inst{4}
\and B. Crosse \inst{2}
\and D.~Davidson \inst{2}
\and A.~DeMarco \inst{11}
\and D.~Emrich  \inst{2}
\and A.~van~Es \inst{14}
\and D.~Fierro \inst{7}
\and A. Faulkner \inst{12}
\and M. Gerbers \inst{1}
\and N. Razavi-Ghods \inst{12}
\and P. Hall \inst{2}
\and L. Horsley  \inst{2}
\and B. Juswardy  \inst{2}
\and D. Kenney  \inst{2}
\and K.~Steele  \inst{2}
\and A.~Magro \inst{11}
\and A.~Mattana \inst{3}
\and B. McKinley \inst{2}
\and J.~Monari \inst{3}
\and G.~Naldi \inst{3}
\and J.~Nanni \inst{9}
\and P.~Di~Ninni \inst{4}
\and F.~Paonessa \inst{8}
\and F.~Perini \inst{3}
\and M.~Poloni  \inst{3}
\and G.~Pupillo \inst{3}
\and S.~Rusticelli \inst{3}
\and M. Schiaffino \inst{3}
\and F. Schillir\`o \inst{6}
\and H. Schnetler \inst{15}
\and R. Singuaroli \inst{10}
\and M. Sokolowski \inst{2}
\and A. Sutinjo  \inst{2}
\and G. Tartarini \inst{9}
\and D. Ung \inst{2}
\and J. G. Bij de Vaate \inst{1}
\and G. Virone \inst{8}
\and M. Walker \inst{2}
\and M. Waterson \inst{14}
\and S. J. Wijnholds \inst{1}
\and A. Williams  \inst{2}
    }

  \institute{
  ASTRON - Dwingeloo, The Netherlands\\ \email{benthem@astron.nl}
  \and ICRAR/Curtin University - Perth, Australia
  \and INAF/IRA - Institute of RadioAstronomy - Bologna, Italy
  \and INAF - Osservatorio Astronomico di Arcetri - Florence, Italy
  \and INAF - Istituto di Astrofisica Spaziale e Fisica Cosmica -  Milan. Italy
  \and INAF - Osservatorio Astronomico di Catania - Catania, Italy
  \and INAF - Headquarter - Rome, Italy
  \and CNR-IEIIT - Turin, Italy
  \and Bologna University DEI - Bologna, Italy
  \and University of Florence DINFO - Florence, Italy
  \and Institute of Space Sciences and Astronomy, University of Malta - Msida, Malta
  \and University of Cambridge - Cambridge, United Kingdom
  \and University of Oxford - Oxford, United Kingdom
  \and SKA Organization HQ - Jodrell Bank, Cheshire, United Kingdom
  \and ATC/STFC - Edinburgh, Scotland
  }
  
     \abstract
    {
    The design and development process for the Square Kilometre Array (SKA) radio telescope's Low Frequency Aperture Array component was progressed during the SKA pre-construction phase by an international consortium, with the goal of meeting requirements for a critical design review.
    As part of the development process a full-sized prototype SKA Low `station' was deployed
 --  the Aperture Array Verification System 1 (AAVS1).
   We provide a system overview and describe the commissioning results of AAVS1, which is a low frequency radio telescope with 256 dual-polarisation log-periodic dipole antennas working as a phased array. A detailed system description is provided, including an in-depth overview of relevant sub-systems, ranging from hardware, firmware, software, calibration, and control sub-systems.
   Early commissioning results cover initial bootstrapping, array calibration, stability testing, beam-forming, and on-sky sensitivity validation. Lessons learned are presented, along with future developments.}


\maketitle
%

\section{Introduction}
The Square Kilometre Array (SKA) radio telescope promises to be a transformational instrument for astronomy and astrophysics in the coming decade \citep{2015aska.confE.174B}. The SKA telescope is split into the Mid-Frequency Array (SKA-Mid) and the Low Frequency Array (SKA-Low), the latter to be constructed at the Murchison Radio-astronomy Observatory (MRO) in Western Australia. It will be the most sensitive instrument ever built in its operational frequency range.

SKA-Low will operate between 50 and 350\,MHz. At these radio frequencies, it is feasible and practical to build very large collecting areas with simple dipole-like antennas operating as a phased array. Many early-generation radio telescopes were constructed with such a design using analogue beam-forming techniques.
Likewise, the core operating concept for SKA-Low converged on the aperture-array design; the telescope consists of hundreds of thousands of fixed antennas arranged in several hundred `stations', where each station acts as one or more phased arrays.
Modern digital electronics allow great flexibility in the beam-forming process. The current design for SKA-Low consists of 512 stations, each of which is composed of 256 antenna elements \citep{SKA1BaselineDesign}.

In the `pre-construction' phase of the SKA, a consortium called the Aperture Array Design and Construction (AADC) Consortium was responsible for progressing the design of SKA-Low up to the formal system engineering critical design Review (CDR) milestone.
Within the high-level overall SKA telescope design, the AADC Consortium was responsible for the Low Frequency Aperture Array (LFAA) part of SKA-Low. This consists of the antennas, analogue signal transport and signal conditioning, digitisation, coarse channelisation, station-level beam-forming, local monitor-and-control, and associated infrastructure \citep{SKA1BaselineDesign}.

The value of building prototype systems as part of the engineering design and development process is well established, especially for large and complex systems containing new and/or unproven technology.
The Aperture Array Verification System (AAVS) was conceived and implemented as a vehicle to support the LFAA engineering design process during the SKA pre-construction phase. The principal motivation for AAVS was to provide a platform to help the design team investigate, mitigate, and retire key risks.
The AAVS programme has eight objectives:
(i) to validate the architecture (as distinct from implementation) and key enabling technologies of the LFAA baseline design; (ii) to assess sub-system prototypes in a realistic operating environment and inform their subsequent engineering development and detailed design;
(iii) to validate key performance requirements and assess, by measurement, progress towards compliance with them; (iv) to gather production knowledge and to get input from large volume production companies to improve the design and reduce costs as well as risks;
(v) to have a test platform to test interfaces regarding timing distribution, data, and control networks;
(vi) to support the development and assessment of LFAA deployment and installation techniques; (vii) to progress development of the tools and techniques -- including correlation, calibration, and digital signal processing -- required to commission and operate SKA-Low; and (viii) to improve the completeness and maturity of the LFAA cost model and budget estimates.
Aperture Array Verification System \#1 (AAVS1) is the name given to the first full-sized SKA-Low station prototype system that contains all components proposed for the LFAA in SKA-Low.

Prior to and during the pre-construction period, several SKA-Low precursors and pathfinders were deployed. These include 
the MWA \citep{2013PASA...30....7T,2018PASA...35...33W},
LOFAR \citep{2013A&A...556A...2V},
HERA \citep{2017PASP..129d5001D}, and
the LWA and LEDA \citep{2013ITAP...61.2540E,2018MNRAS.478.4193P}.
In addition, the Engineering Development Array \citep[EDA1;][]{2017PASA...34...34W} was initially deployed within the context of the SKA cost control process as a potential hybrid signal processing architecture system.

The precursors and pathfinders share many key science goals with SKA-Low and provided valuable on-the-ground and on-the-sky expertise to help inform the SKA-Low design process. Several SKA system-level requirements were updated or created in response to the lessons from the precursors and pathfinders, in particular emphasising the importance of instrumental effects in very high dynamic range experiments. The detection of the Epoch of Reionisation power spectrum -- the SKA's highest priority science goal -- is a prime example of such a high dynamic range experiment.

This paper presents the context in which AAVS1 was developed, the design of AAVS1, the expertise gained from deploying and commissioning AAVS1, and the results of the commissioning process. 

\section{Background}

The need for an SKA LFAA demonstrator of a scale comparable to one SKA-Low station was understood to be a CDR imperative from at least the time of the SKA preparatory programme (PrepSKA), which commenced in 2008.  With a track record of designing and commissioning instruments such as LOFAR and MWA, it was clear to proponents of SKA aperture arrays that AAVS1 needed to be a substantial undertaking. A verification system that retired risk to the satisfaction of the SKA design authority (regardless of any external requirements) needed to be both on site and functional as an astronomical imaging and metrology tool; “On-site, on-sky” was always the goal.
The verification programme was an ambitious one for the AADC Consortium and, in practice, represented the most ambitious of the SKA pre-construction projects.
To realise AAVS1's goals meant that a full range of allied systems needed to be provided in order to produce an astronomically capable verification platform.

Two main strategies were adopted to progressively retire AAVS1 technology and deployment risks.  First, smaller scale verification systems were built in the UK and on site in Australia.  These prototypes, known as AAVS0 and AAVS\,0.5, respectively, were valuable and produced substantial SKA inputs in their own right.
Indeed, AAVS\,0.5 was used to make the first interferometric SKA antenna images \citep{7293140}, as well as show the initial beam pattern measurements using a transmitter on an uncrewed aerial vehicle \citep{2018ExA....10..1007}.
In a second strategy, an alliance between the AADC Consortium and the MWA Consortium was formed, allowing AAVS 0.5 and AAVS1 to use the site infrastructure and network provided by the MWA, and to gain considerable calibration leverage from the adjacent synthesis telescope.
While essential to achieve AAVS1 goals, on-site establishment of the instrument required the cooperation of a number of bodies, including CSIRO, MWA, the SKA office and Curtin University; in these negotiations the AAVS1 Basic Specifications \citep{AAVS1Design,2016icea.confE...1H} were invaluable.


The LFAA will be a discrete sub-system within the full SKA-Low. It has a variety of external dependences (interfaces) that mean it will not provide a ‘telescope’ capability except in the context of SKA-Low. In order to serve as an effective risk mitigation platform, the AAVS1 must be capable of operating in the absence of the external contributions that other elements of the SKA-Low system will provide for LFAA. AAVS1 achieves this by relying on existing MRO infrastructure, provided via the MWA, and the use of limited-function emulators.

AAVS1 was entirely paid for by the membership of the AADC Consortium. Cost considerations had significant influence on the system. Key details of the implementation that were influenced by cost include:
    (i) \textbf{Overall scope:} The overall scope of AAVS1 was limited by the funding available within the Consortium;
    (ii) \textbf{Site preparation:} The extent of civil works possible to establish the AAVS1 was limited by the funds available. The limited ‘scrape to clear’ approach taken in AAVS1 is an excellent test of whether clearing the ground to the prevailing terrain profile is sufficient for LFAA,
    as opposed to a more complicated and costly option of preparing and levelling the ground to some specification of flatness;
    (iii) \textbf{Fibre installation:} The budget did not allow for the ~5000\,m optical‐fibre trunk cable that connects AAVS1 to the MRO control building to be buried. Having this cable surface laid and therefore exposed to the environment represents a ‘worst case’ scenario for SKA-Low in terms of temperature variations. AAVS1 provides an excellent opportunity to evaluate the feasibility of this implementation should it become necessary for any reason.
    (iv) \textbf{Representative components:} In many areas AAVS1 makes use of functionally representative commercial off-the-shelf (COTS) components, whereas LFAA will utilise custom specified items that would be prohibitively expensive in smaller quantities. The benefits and/or difficulties experienced using COTS components in AAVS1 can help to inform the custom specification for LFAA, or even demonstrate the feasibility/utility of using the COTS components for LFAA.

The construction of AAVS1 started early 2016, by deploying several single antennas at the AAVS1 site, to ensure full signal-chain functionality and help assess field-deployment issues. Most of the second half of 2016 was used to produce and factory-test all hardware within Europe, prior to shipping to the MRO.
In parallel, the station locations were prepared at the MRO, providing station infrastructure and deploying ground planes, concrete antenna bases, and antenna and power interface units (APIUs). 
In March 2017, an initial set of 40 antennas were deployed and tested at the MRO, resulting in a first station-level test setup. Unfortunately initial results showed technical issues, including an auto-oscillation of some of the antennas, as described in Sect. \ref{sec:testandverification}. During a long field trip in November 2017, remediation work to eliminate the oscillation issue was performed, resulting in a fully deployed AAVS1 station. Subsequent work to bootstrap the system and verify the signal chain was performed remotely.

\section{The AAVS1 system}
\label{sec:signal_path}

\begin{figure*}
	\includegraphics[width=\textwidth]{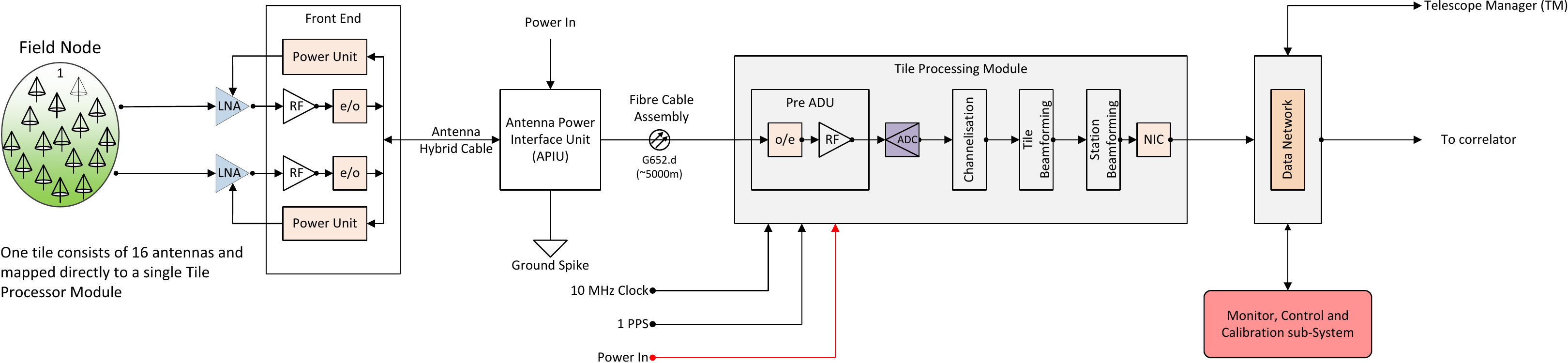}
    \caption{AAVS1 system signal chain block diagram. The fibre cable assembly in the centre of the image represents the 5.2\,km of optical fibre connecting the in-field components to the MRO central control building. The components to the left of the fibre cable assembly are in the field, components to the right are in the control building.}
    \label{fig:AAVS1_blockdiagram}
\end{figure*}

AAVS1 is a prototype realisation of the LFAA including antennas, analogue signal conditioning and transport, signal digitisation and channelisation, beam-forming, power distribution, synchronisation, and local monitor and control. A block diagram of the system signal architecture is shown in Fig. \ref{fig:AAVS1_blockdiagram}. A high-level overview is provided here with detailed sub-system descriptions in the following sections.

\begin{figure}
	\includegraphics[width=\columnwidth]{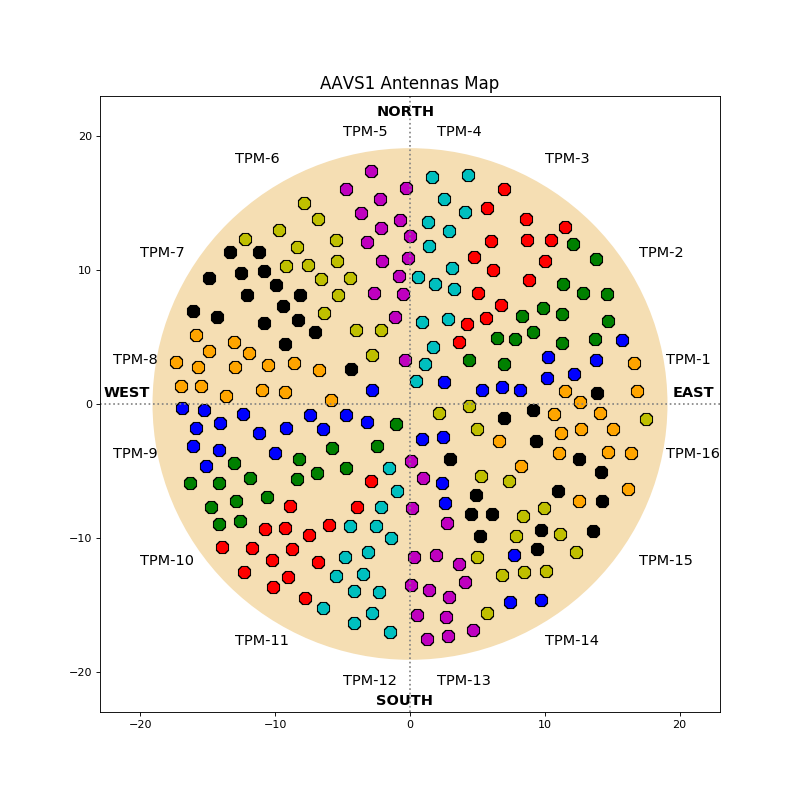}
    \caption{AAVS1 layout. Distances are in metres. Antennas are colour-coded by the TPM they are connected to in sets of 16, with each set roughly forming a wedge in the station. The APIU (not shown) resides in the centre of the station.}
    \label{fig:AAVS1_layout}
\end{figure}

The AAVS1 station comprises 256 SKALA2 antennas mounted on a coarse-pitch wire ground screen. The antenna layout is a pseudo-random distribution within a 35\,m diameter circular footprint (see Fig. \ref{fig:AAVS1_layout}). 

An AAVS1 station is composed of the following four parts:
The first are the 256 dual-polarisation SKALA2 antennas, each including low-noise amplifiers (LNAs) and optical front-end module (FEM), which converts electrical radio frequency (RF) signals to optical signals.
The second is the  APIU, which delivers power to, and receives signals from, the antennas. The antenna is connected to the APIU by a custom hybrid cable that carries power to the antenna and receives the optical signals coming from the antenna. The APIU provides the local optical fibre patch panel for connecting the optical signals from each antenna to the 5.2 km long optical cables that go to the MRO control building.
The third are the tile processing modules (TPMs), where fibres from the field terminate in the control building. Each TPM accepts 16 fibre inputs hence services 16 individual antennas (32 signal paths). Tile processing modules contain both analogue and digital components as described below.
The fourth is the monitor calibration and control sub-system (MCCS), which controls sub-systems, provides calibration systems and additional signal processing and computing functionality.

\subsection{Antennas and analogue modules}
\begin{figure}
	\includegraphics[width=\columnwidth]{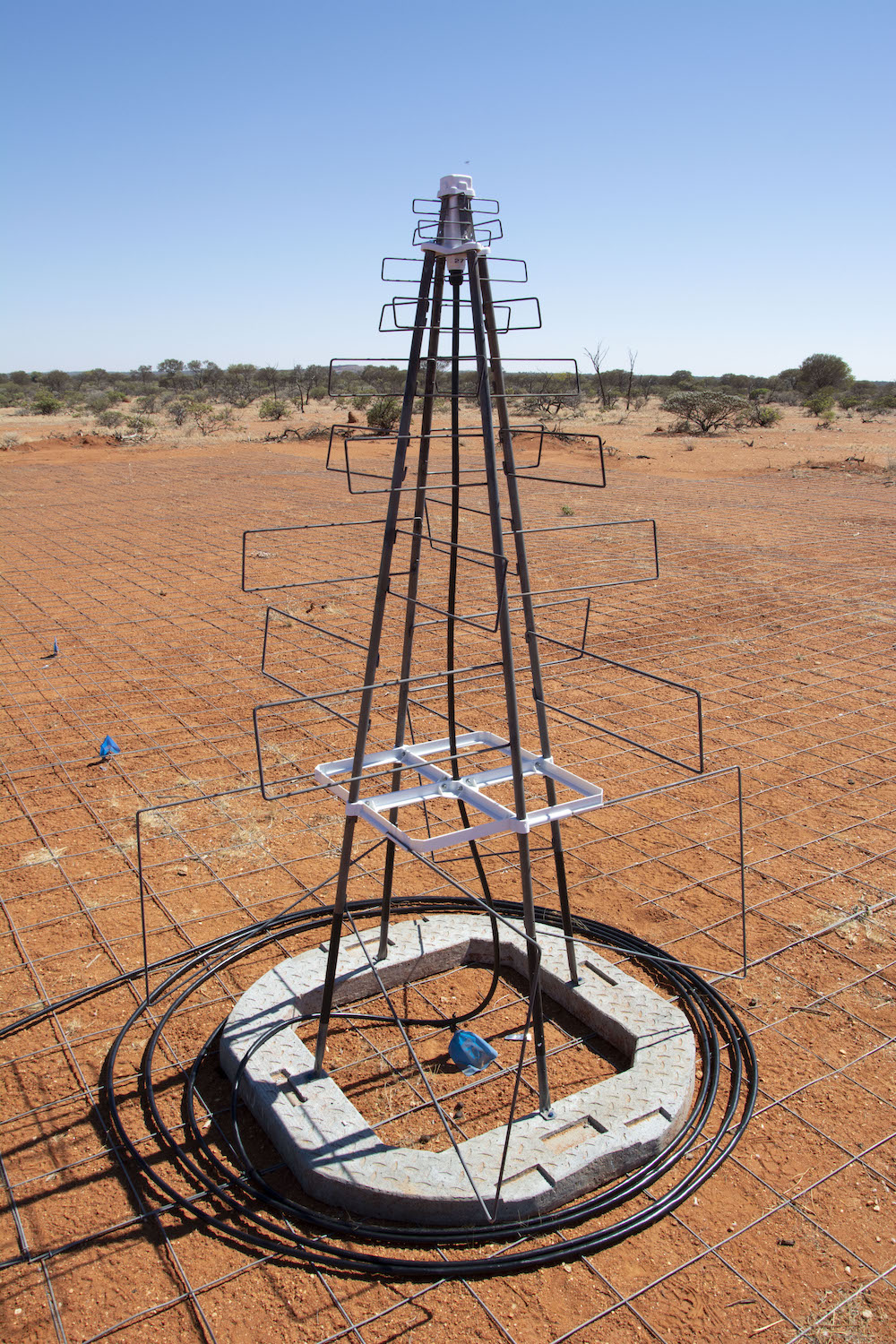}
    \caption{SKALA2 antenna installed as part of AAVS1.}
    \label{fig:SKALA2}
\end{figure}

A SKALA2 wideband dual-polarised log-periodic antenna is shown in Fig. \ref{fig:SKALA2}.
We summarise its main features here and refer readers to \citet{2015ExA....39..567D} and \citet{7297231} for details.
Each antenna is mounted on a concrete base that sits atop the wire mesh.

The array configuration was made to be pseudo-random while maintaining roughly uniform antenna density and avoiding overlapping antennas.
Pseudo-random configurations have been used in other radio telescopes that have large fractional bandwidth (e.g. LOFAR, LWA), and once again in SKA-Low in order to avoid grating lobes and scan blindness, at the expense of an increase in the sidelobe level. It has been demonstrated as well how the effects of mutual coupling randomise out for this type of configuration \citep{7731490}.
The configuration and spacing between antennas was set to optimise it for sidelobe level performance\citep{6623095}. The average spacing between antennas is approximately 1.9\,m.

\subsubsection{LNA}
The LNA is located at the feeding point of the SKALA2 antenna for optimum noise performance. The LNA, for each polarisation, consists of a pair of identical low-noise transistors directly feeding each arm of the antenna and are closely followed by a wideband balun to provide a single-ended output. The differential input has been designed to provide an optimal noise match to the antenna impedance. The in-band gain is about 40\,dB which drops abruptly below 70\,MHz due to the mismatch loss between antenna and LNA. This high-pass response helps to attenuate the strong radio frequency interference (RFI), which can be found up to approximately 30\,MHz.

\begin{figure}
	\includegraphics[width=\columnwidth]{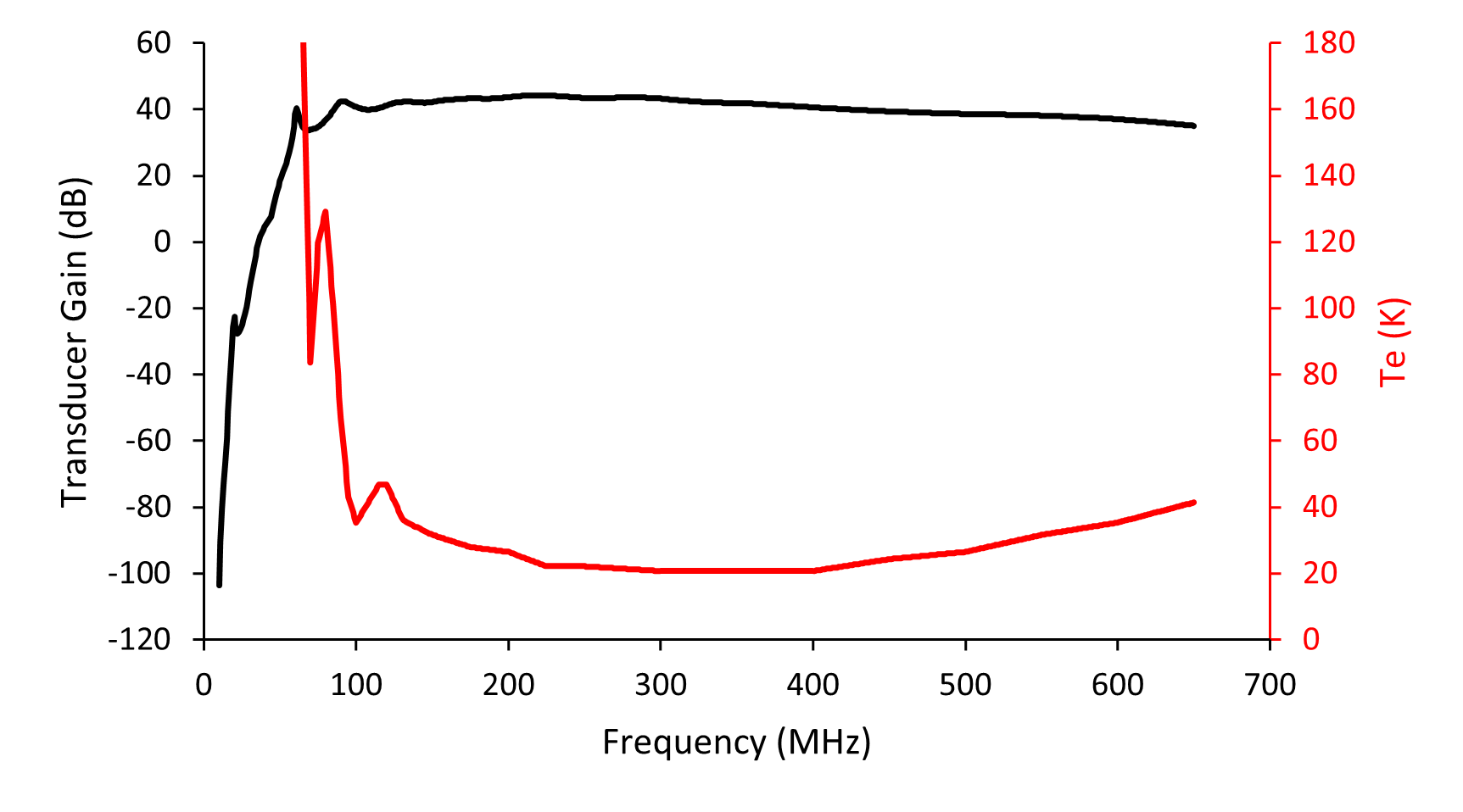}
    \caption{SKALA2 LNA transducer gain (blue) and noise temperature (red).}
    \label{fig:LNA transducer gain}
\end{figure}

As shown in Fig. \ref{fig:LNA transducer gain}, the SKALA2 LNA provides approximately 40\,dB gain and a noise temperature better than 40\,K across most of the frequency range of interest. The plot is a combination of measurements (using antenna measured input reflection coefficient, S11) and simulations of the LNA\footnote{Earlier versions of the SKA-Low requirements called for the telescope to work to 650\,MHz.}.
The SKA model sky temperature \citep{SKA1L1req} is approximately $T_{sky} = 2.73+20*(408/\nu_{MHz})^{2.75} K$, which equals the LNA noise temperature around 425\,MHz and is well above the LNA temperature below 250\,MHz.

\subsubsection{Optical front-end module} 
The optical FEM is connected to the LNAs by two short coaxial cables. The RF signals, before they are converted to the optical domain, are further amplified by about 20\,dB in order to reduce the added noise of the entire receiver chain to the system temperature as much as possible. The FEM also filters out strong RFI below 50\,MHz and provides power to the LNAs of both polarisations via a bias tee.

The optical transmitter is based on the simple direct intensity modulation of a low-cost, high-performance distributed feedback (DFB) laser. This approach is feasible due to the relatively low radio frequencies in the system.
One single optical fibre can transfer the two polarisations of each antenna thanks to the wavelength division multiplexing (WDM) technique, where each RF signal is carried by a different optical wavelength (1270 and 1330\,nm).

\begin{figure}
	\includegraphics[width=\columnwidth]{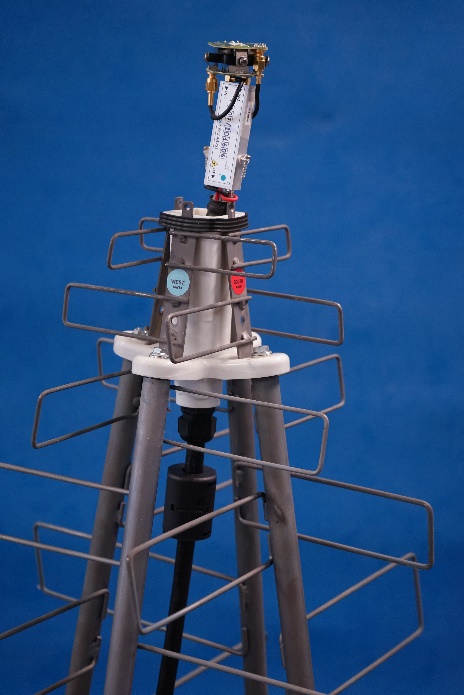}
    \caption{Antenna electronics showing the top of a SKALA2 and the contents of the trumpet, including LNAs and the FEM. The hybrid cable runs down the centre of the antenna.}
    \label{fig:Antenna electronics}
\end{figure}

Both LNAs and the FEM are encapsulated into a custom plastic enclosure referred to as `trumpet', which constitutes the apex part of the SKALA2 antenna. The signals from each antenna are then routed to the APIU through an optical pigtail enclosed in the hybrid cable as shown in Fig. \ref{fig:Antenna electronics}.
The length of the hybrid cable (23\,m) allows every antenna to reach the APIU regardless its position in the station. For the antennas closest to the APIU, the extra cable length is coiled around their respective base. For AAVS1, all hybrid cables were the same length, although the delay compensation capability in the TPMs can correct for non-equal lengths.

The nominal power budget for the station was 802\,W into the APIU, of which nominally each antenna consumed 2\,W. The APIU contains no monitoring capability, however, so the actual power consumption was not measured.

\subsection{The 5.2 km fibre optic cable}
The RF modulated optical signals from each antenna are transported via a multi-core fibre optic cable (with 576 cores) to the MRO control building, around 5.2 km away from the location of AAVS1.
The fibre optic core complies with the ITU-T G.652-D standard\footnote{\url{https://www.itu.int/rec/T-REC-G.652-201611-I/en}}. As the zero dispersion wavelength of the core is at 1310 nm, to minimise the RF phase variation induced by chromatic dispersion of the optical media \citep{Juswardy2015}, RF signals from two different polarisations in an antenna are transmitted on two distinct wavelengths as described in the previous section.

The construction of the cable consists of eight units of 6 x 12 fibre called spider web ribbons (SWRs)\footnote{Non-Metallic Armoured SWR Fibre Optic Wrapping Tube Cable, product code: A57SLF8576BK, AFL Global}, which makes the overall diameter of the cable assembly much smaller than a conventional loose-tube cable. This is desirable to minimise the temperature difference of the fibre cores across the cross-section of the cable, as the temperature difference across the fibre cores is directly associated with the RF phase stability of each receiver link \citep{Juswardy2015b} due to thermal expansion, as well as change in refractive index, along the 5.2\,km cable.

\begin{figure}
	\includegraphics[width=\columnwidth]{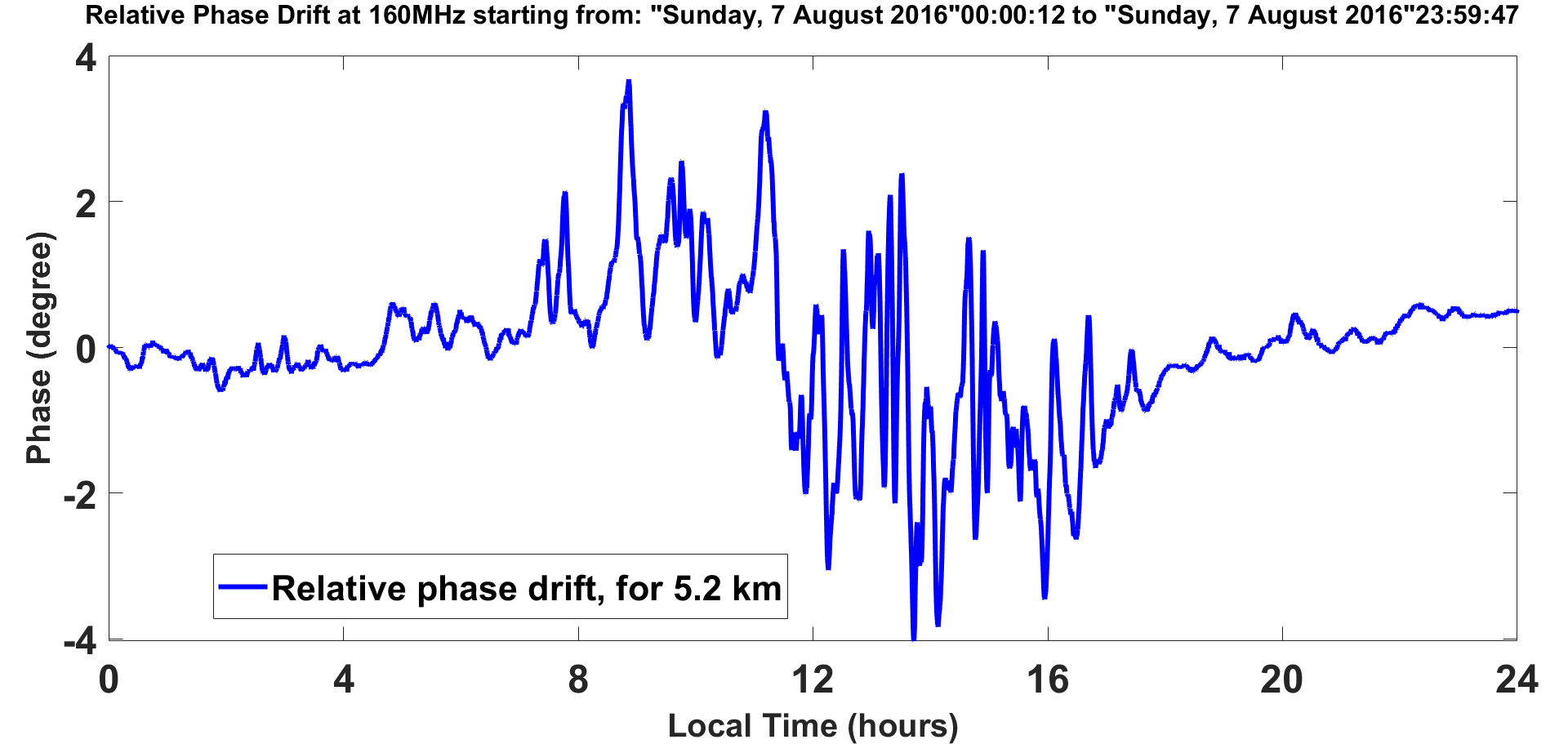}
    \caption{Relative phase difference at 160 MHz of the AAVS1 long fibre cable. Shown is the result for the case where different transmission wavelengths were used, with each wavelength transmitted on a different fibre core selected from opposite locations inside the cross-section of the cable.}
    \label{fig:fibre_cable_phase}
\end{figure}

The relative RF phase variations in the SWR cable due to environmental exposure in the field have been quantified through a series of measurements at the MRO. A snapshot of the measurement results is shown in Fig. \ref{fig:fibre_cable_phase}, where the 24-hour relative phase difference from two different wavelengths was taken.
Each wavelength was transmitted on a different fibre core, in opposite locations inside the cross-section of the cable. This was in order to simulate a possible worst-case scenario of the relative phase difference from two separate antennas of disparate polarisations.

These results can be used to estimate the performance of future fibre cable installations for any given length, and to inform decisions on, for example: the calibration timescale; the maximum cable length or transmission distance; whether underground cable trenching is required; or the selection of a suitable cable assembly (standard loose-tube or SWR cable).
For AAVS1, the small phase variations between fibres of the surface-laid cable were deemed to have negligible effect on performance.

\subsubsection{PreADU}
The analogue part of the TPM is the `preADU'. 
The preADU receives the optical signals via fibre input. A WDM optical receiver first separates the two optical wavelengths and then converts the signals back to electrical through their direct detection with two PIN photodiodes.
The signals are then further amplified, and, using a digital step attenuator (DSA), it is possible to adjust their amplitude for optimal digitisation. Particular care in the optical WDM components, and in the FEM and preADU printed circuit board (PCB) board design, has been taken in order to meet the RF isolation specifications between the two RF channels (ISO>30\,dB).
In addition, preADU components (Fig \ref{fig:iTPM}) are encased in shielded boxes to prevent external crosstalk or interference from digital sub-systems.

Two more functions have been included in the preADU receivers: a $50 \Omega$ switchable load and a filter bank. The $50 \Omega$ load is intended for debugging purposes at the receiver level since no remote power control of the antenna electronics in the field is provided in AAVS1.
The filter bank is composed of a low-pass filter at 375\,MHz and a band-pass filter at 375-650\,MHz. The former has to be used with a 800\,MHz clock (first Nyquist zone), while the latter requires a 700\,MHz clock (second Nyquist zone) and was introduced to allow the option to work beyond the SKA-Low band.

\subsubsection{Overall receiver behaviour} 

\begin{figure}
	\includegraphics[width=\columnwidth]{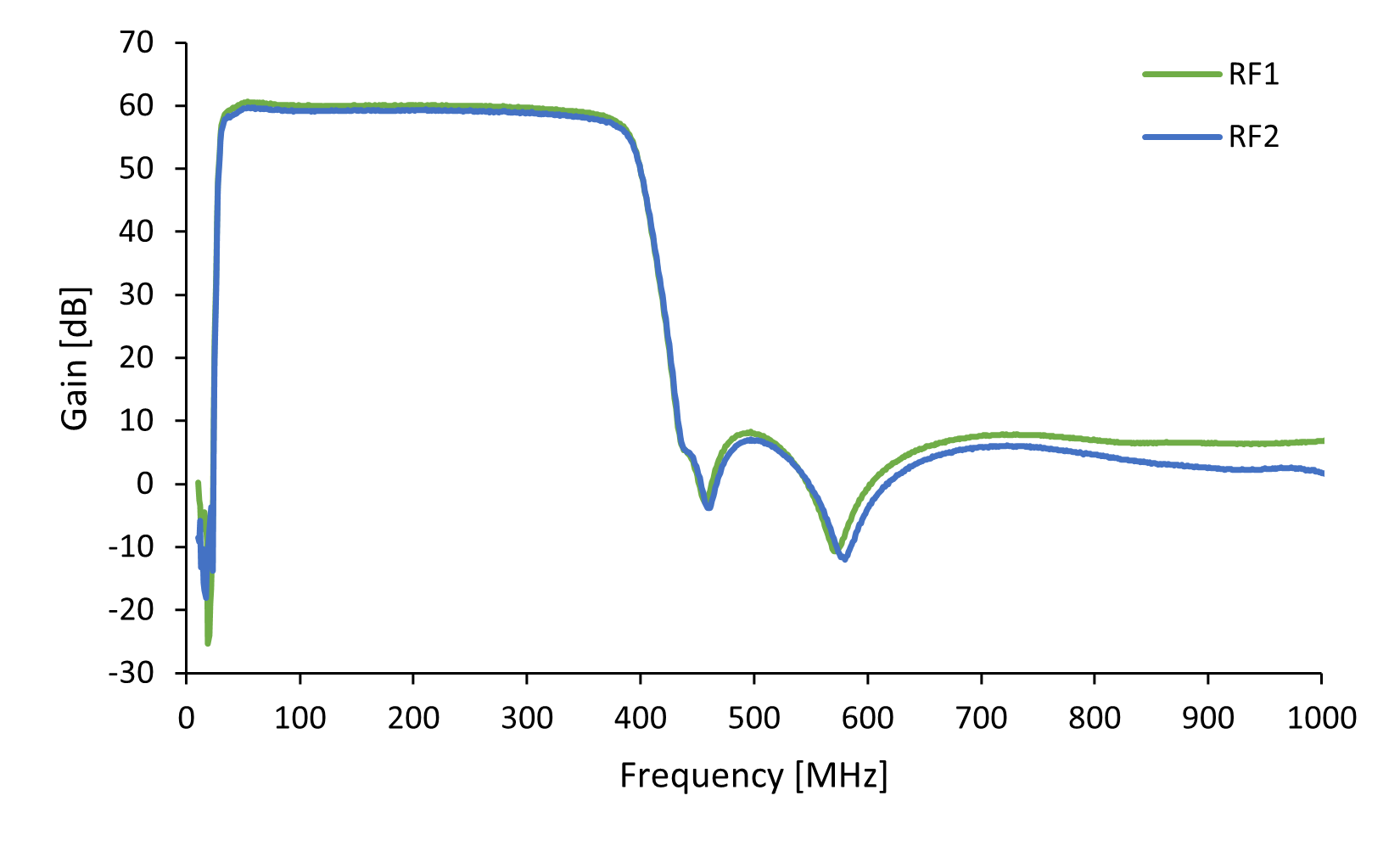}
    \caption{Analogue system gain from the input to the FEM, DSA at 0 dB attenuation.}
    \label{fig:Receiver gain}
\end{figure}

The DSA is used to statically compensate for different signal levels due to the spread of the gain among individual analogue chain components, such as the LNA, FEM, and preADU, and for the differences in the attenuation of the optical fibres.
For these reasons, in order to have a robust design, the receiver provides the nominal gain of 41dB when the DSA is set to its mean value of 15\,dB (the DSA range is 0-31\,dB with 1\,dB steps) and when the FEM is connected to the preADU through the 5\,km optical fibre cable, which introduces 2.2\,dBo of optical attenuation (equivalent to 4.4\,dB RF attenuation).
Figure \ref{fig:Receiver gain} shows the total gain of the analogue system from the input of the FEM to output of the preADU with 0\,dB attenuation in the DSA.


\subsection{Digital signal path}
\label{sec:digital_path}
\begin{figure}
	\includegraphics[width=\columnwidth]{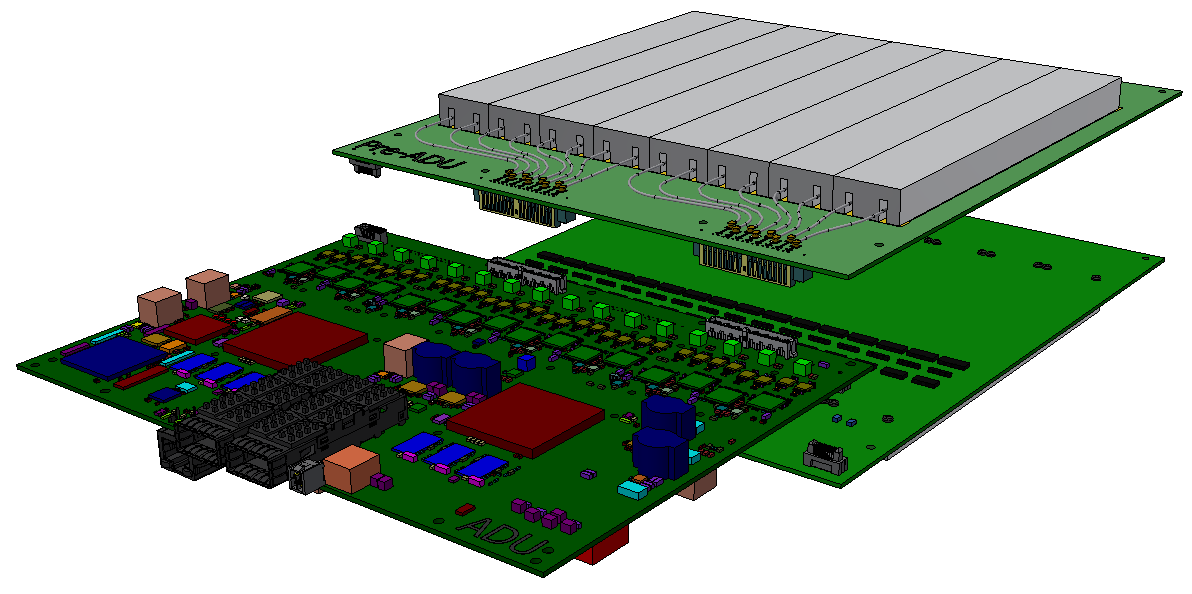}
    \caption{TPM ver.1.2 design (iTPM). Right: Two preADU boards with eight optical receivers each. Left: Analogue to digital unit.}
    \label{fig:iTPM}
\end{figure}

The AAVS1 digital signal processing hardware is hosted in four racks in the MRO control building.
The infrastructure provides an air cooling system that guarantees a constant temperature of 18 degrees Celsius within each 42U cabinet, and high-accuracy time and frequency references instantiated by a maser atomic clock. Off-the-shelf components distribute the synchronisation signals, network and power to each TPM.
Five Ettus Research Octoclock devices deployed in a tree topology split the incoming synchronisation signals (10\,MHz reference and Pulse Per Second) to serve the 8 TPMs of each rack placed in two 14U 19\,-inch wide sub-racks. They host four 6U TPM boards and, using fans, force cold airflow from the front of the rack to the board's heat sinks.
The TPMs are composed of two identical preADU boards and a main processing unit called the ADU (`analogue to digital unit'), shown in Fig. \ref{fig:iTPM}.

Two private network infrastructures support the TPM's monitor and control functions (1\,Gb Netgear Switches) and the calibration and beam-forming traffic (40\,Gb Mellanox Switches).
The two networks are redundantly connected to each other using 4x10\,Gb split cables for fail-safe recovery. Passive copper cables connect devices within the rack, while optical active cables are used for rack to rack connections.


\begin{figure}
	\includegraphics[width=\columnwidth]{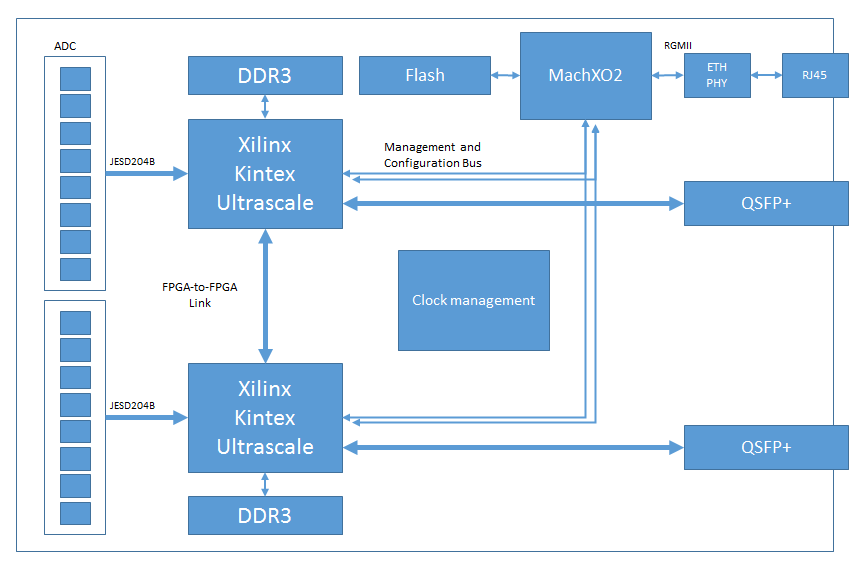}
    \caption{ADU block diagram.}
    \label{fig:iTPM_block_diagram}
\end{figure}

The ADU is the main component of the digital signal processing path \citep{2017JAI_HW}, as shown in the block diagram in Fig. \ref{fig:iTPM_block_diagram}.
The ADU is a 6U board that hosts two Xilinx XCKU040 field programmable gate arrays (FPGAs) and sixteen dual-input AD9680 analogue to digital converters (ADCs), supporting 14 bit sampling and up to 1 GSPS sampling rate; however, in AAVS1 only the eight most significant bits are used. Each ADC digitises two polarisations from a single antenna at 800 MSPS and transmits the encoded digital data stream to the FPGA over two JESD204B data lanes running at nominal rate of 8 Gbit/s. 

Each FPGA supports onboard 64-bit-wide DDR3 memory consisting of four DDR3 memory chips for a total capacity of 2\,GByte per ADU. High-speed data communication is enabled by a dual QSFP connector supporting one 40\,GbE link per FPGA. In order to  transfer data locally between the FPGAs, a high-speed bidirectional LVDS bus has been implemented.

The TPMs within the AAVS1 station are synchronised by an externally provided 10\,MHz reference clock and pulse-per-second (PPS) signal.
On each ADU an AD9528 phased-locked loop (PLL) provides coherent 800\,MHz and 100\,MHz clocks, which are phase-locked to the reference clock. The first generated clock is distributed to the ADCs and used as sampling clock, while the second is used in the FPGAs as a reference clock for the JESD204B transceiver. 
The PPS is used as trigger signal to synchronise the ADUs within the station.

The management and control of the board is performed by a dedicated functional block which includes a Lattice MACH-XO2 complex programmable logic device (CPLD) that manages a gigabit ethernet link with RJ45 interface. The CPLD implements a firmware UDP/IP stack and it is able to decode the protocol used by MCCS to control and monitor the ADUs. Using universal computer protocol (UCP) packets, MCCS can instruct the CPLD to perform different tasks such as reading and writing registers in the FPGAs, accessing ADC internal registers, or reading the onboard temperature sensor through I2C.

The firmware for the Xilinx FPGAs is described in \cite{2017JAI.....641015C}. It is implemented in VHSIC Hardware Description Language (VHDL) and is based on a highly modular structure that separates the technology-dependent part of the firmware, included in an input-output (IO) module, from the actual digital signal processing (DSP) design. The IO module implements all the board interfaces, such as DDR3, ethernet, and F2F, that are specific to the ADU board. 
This separation facilitates the maintenance of different DSP designs within the same VHDL code base and will aid in the process of transitioning to future hardware platforms. The ADU boards and related code base have been successfully used or are planned to be used in several instruments \citep{pharos2,mexart}.




The AAVS1 digital signal processing structure is based on a frequency domain beam-forming architecture. 
Signals from the individual antennas are channelised into 512 channels with a spacing of 781.25\, kHz, and finer delays can be applied in the frequency domain by applying a dynamic phase correction to each frequency channel. Receiver and atmospheric gain variations can be corrected by applying a complex calibration factor for each frequency channel.
Delay can be specified either as the phase of this factor or directly as a delay, with the firmware computing the appropriate phase factor for each frequency channel. A delay rate can also be specified for sidereal tracking.

The channeliser uses an over-sampled time-multiplexed polyphase filter-bank structure, and is part of a two-stage channelisation scheme. Oversampling by a factor of 32/27 allows the subsequent correlation to produce a continuous seamless cross spectrum by discarding the edges of each channel and keeping the central part, with a flat spectral response. 
The filter response is flat to $\pm 0.2$\, dB in the pass band and has at least 60\, dB, typically 90\, dB of out-of-band rejection.
The filter is based on the Remez-McClellan algorithm, using a fraction of the filter taps for convergence, and is interpolated to the final number of taps \citep{Comoretto2012}.


The firmware includes a set of test features. A test generator can generate pseudo-random white noise, with adjustable fixed delay, and up to two sinusoidal tones. The signal at various stages of the processing can be buffered, formatted into frames, and sent on both the 1\,Gb and 40\,Gb ethernet links. Total power data can be computed, both for the broadband signals and for the channelised data, in this case effectively producing
a coarse resolution spectrum. These capabilities are further described in Sect. \ref{sec_commissioning}.

The firmware implementation has been highly optimised for the Xilinx Ultrascale FPGA family by using a set of low-level VHDL modules tailored to the FPGA architecture, but is otherwise written in a very generic way. 
This eases porting the high-level functionalities to different architectures, but at the same time allowed to fit 16 parallel processing chains in one relatively small FPGA and reduce the TPM power requirements. 
The optimised channeliser uses 62 multipliers per signal and the design occupies about 85\% of the FPGA resources.

\subsection{Monitor calibration and control sub-system}
\label{sec:MCCS}
AAVS1 includes a single processing server that performs all the software operations for the array which can be grouped into two primary services: hardware monitoring and control; and real-time and offline data processing. The MCCS is also able to provide a test platform towards the interface of the Telescope Manager Consortium. 

A hardware monitoring and control module, PyFABIL \citep{pyfabil}, was developed to communicate with TPMs and other digital boards. It presents a uniform layer through which hardware devices can
be accessed. For the TPM, UCP is used for communication, while XML is used to  describe the registers and memory address spaces of the CPLD and firmware running on the two FPGAs. This XML file is generated automatically during bitstream compilation and then concatenated to the bitstream itself.
The register and memory map is then retrieved by a client upon connection, such that the package can reflect the functionality available in the firmware.
When firmware is loaded onto FPGAs the functionality of the board
changes and firmware-specific initialisation, synchronisation, checks, and monitoring and control have to be performed.
PyFABIL employs a plug-in-based system that can be dynamically loaded to a board instance during runtime. Plug-ins were developed for all the major hardware components on the board (board-specific plug-ins) as well as the major digital processing blocks in the AAVS1 firmware (DSP-specific plug-ins).
Higher-level software components then use PyFABIL to perform station-level operations, such as synchronising the TPMs
together, forming the beam-forming chain, issuing synchronised commands (for example, for data transmission) and equalising the RF chains. 

A high-performance data acquisition library was developed to receive  streaming data from the TPMs \citep{daq}. This is composed of a low-overhead raw-frame receiver that is attached to a network interface and a number of "packet processors", here referred to as consumers, which can interpret and process the incoming packets. A consumer must register a packet
filter with the frame receiver such that it only forwards frames of interest to the consumer. All consumers write the received
data to disk in HDF5 format\footnote{https://www.hdfgroup.org/} through an interfacing Python layer.

The TPM firmware can transmit a variety of data stream types, the most useful of which are: 1) channelised data from each antenna, which is used for intra-station correlation,
2) integrated channelised data from each antenna, which is used for basic station health and bandpass monitoring,
and 3) station data, the main data product of the TPM, which transmits the calibrated, beam-formed station beams to downstream processors (see Sect. \ref{sec_commissioning}).
In total, seven SPEAD\footnote{Streaming Protocol for Exchanging Astronomical Data, https://casper.ssl.berkeley.edu/wiki/SPEAD} data stream types can be transmitted, each with a different data rate and packet content structure.
An item in the SPEAD header specifies which data type the packet contains, such that it can be used to route the packet to the appropriate consumer. 

\subsection{Capabilities available for commissioning}
\label{sec_commissioning}

The AAVS1 firmware provides a number of data taps in between digital processing blocks for transmitting intermediary data streams that are useful for debugging and commissioning. This goes in hand with the software data acquisition capabilities. These are referred to as acquisition modes, and are summarised below.

\begin{description}
\item \textbf{Raw voltages}: A snapshot of the raw digitised voltages (1.024\,$\mu$s or 8k samples in synchronised mode and 81.92\,$\mu$s, or 32k samples, otherwise) is transmitted before the channelising stage. This is useful for checking for ADC saturation, clipping and general input signal characteristics.

\item \textbf{Channelised voltages}: Raw channelised voltages (the output of the channeliser) are transmitted. By default, frequency channels are transmitted in a round-robin fashion, starting from channel 0 up to channel 511, with a selectable duration per channel. An extension of this mode is to transmit continuously one channel of interest. The channelised data are transmitted synchronously across all TPMs, generating approximately 8\,Gbps. Due to this data rate, the data stream cannot be stored to disk in real time, so an intermittent mode is also available, which is capable of storing segments (approximately 1\,s) of data at a specific cadence. Channelised  data can be correlated in real time (see Correlater mode), correlated offline, or beam-formed offline for further processing.

\item \textbf{Integrated channels}: The AAVS1 firmware includes an integrator that can generate integrated total power spectra for all antennas and transmit these at a low data rate. This is primarily used for constantly monitoring the bandpass of all the antennas.

\item \textbf{Correlator mode}: In correlator mode, xGPU \citep{xGPU} is attached to the channelised data mode above to correlate the incoming data stream in real time. This can be used in either continuous mode (single channel) or burst mode (all channels in round-robin), generating a correlation matrix at $\approx$1.98\,sec time resolution. Correlator mode is used within a calibration service, which automatically generates calibration solutions for all antennas once a new correlation matrix is generated. This service was typically run once a day at the time of solar transit in order to use the Sun as a point source (reasonably good approximation for the 35\,m diameter of the station) calibrator (Sect. \ref{sec_calibration}).

\item \textbf{Station beam}: The main products of the station are station beams, which in SKA-Low will be transported to Central Signal Processing (CSP) for real-time correlation with the rest of the array, with a data rate of \texttt{\char`\~}20\,Gbps per full band beam. For commissioning, this data stream is integrated in real time and stored to disk at a reduced data rate. 
\end{description}

In order to validate the real-time beam-forming process, the real-time station beam and channelised voltages were collected simultaneously at the same frequency channel. It was verified that the total beam power over 24\,hours obtained from the real-time beam-forming procedure was exactly the same as the offline beam-formed channelised voltages and compared against the simulations (Sect. \ref{sec_beamforming}).

\begin{figure}
	\includegraphics[width=\columnwidth]{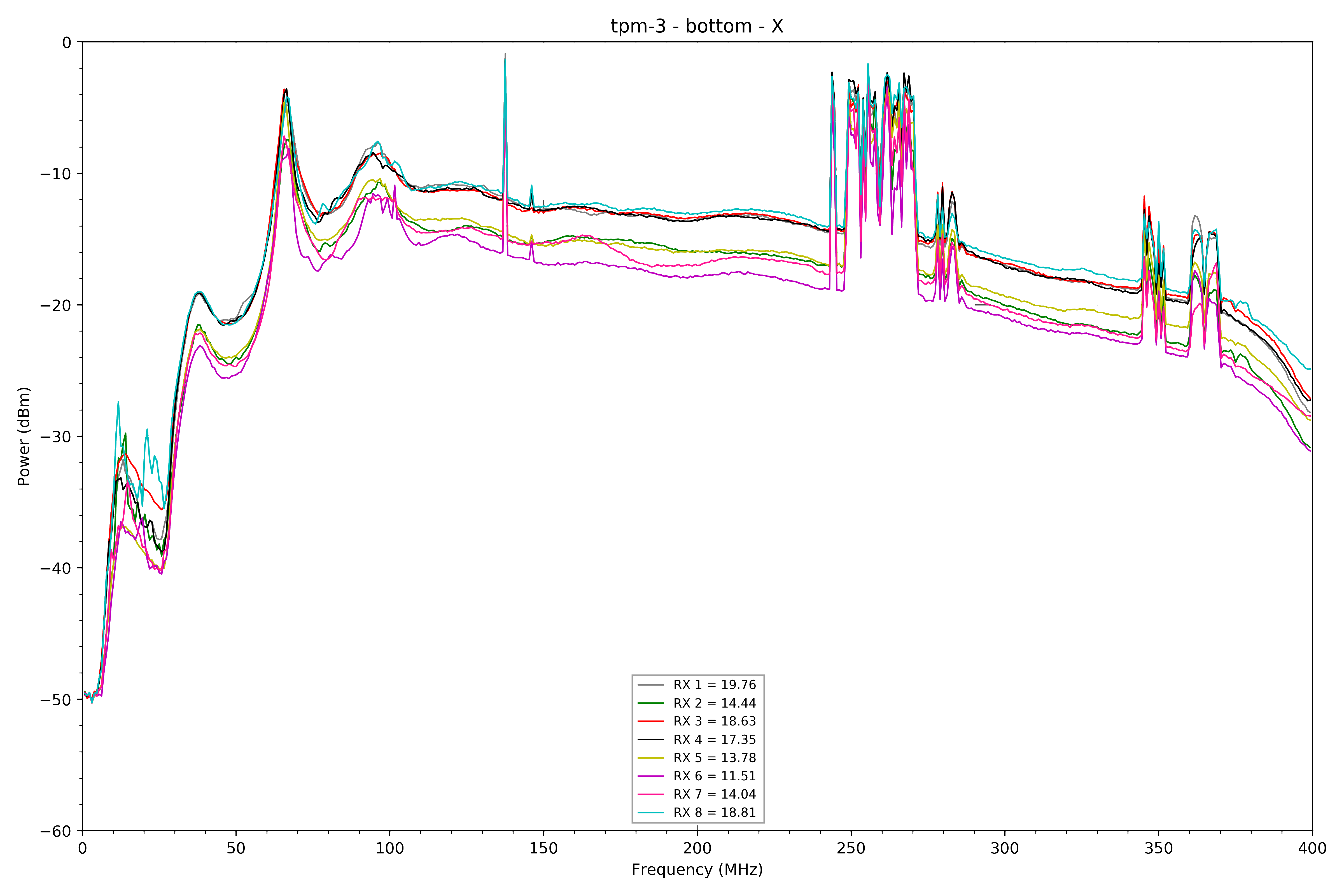}
    \caption{Example uncalibrated integrated channel total power spectra from a TPM. This is indicative of the typical signal generated by the SKALA2 antennas. Narrow signals at 137\,MHz and at higher frequencies (245--285\, MHz and above 340\,MHz) are due to known satellite-based RFI sources. The Y polarisation (not shown) is qualitatively equivalent.}
    \label{fig:AAVS1_example_spectra}
\end{figure}

\begin{figure*}
	\includegraphics[width=\textwidth]{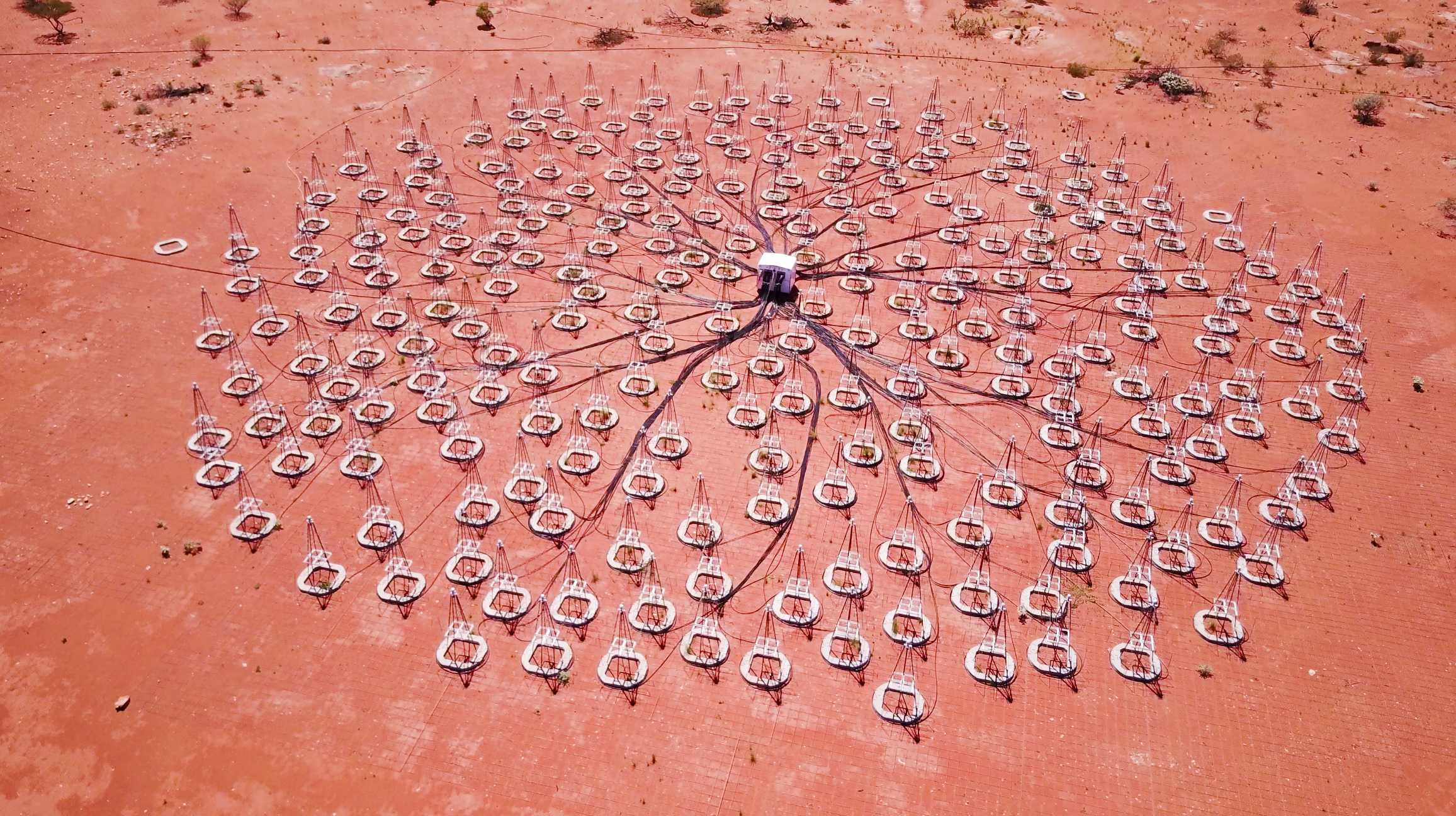}
    \caption{AAVS1 in February 2018. The white box in the centre is the APIU, with the black hybrid cables running from each antenna to the APIU.}
    \label{fig:AAVS1_aerial}
\end{figure*}

Example total power spectra from eight antennas on one of the TPMs is shown in Fig. \ref{fig:AAVS1_example_spectra}.
The fully deployed and operational AAVS1 station is shown in Fig. \ref{fig:AAVS1_aerial}. Visible are the antennas, APIU, hybrid cables and multi-core 5\,km fibre optic trunk cable heading off to the MRO control building.

\section{Array calibration using interferometry}
\label{sec_calibration}

While the total power plots from each antenna allow the gain to be adjusted for each antenna's amplitude, the differential delays between antennas are unknown after the station is turned on. In addition, it is important to confirm the signal mapping within the system from polarisation on the  antenna all the way through to the digital signal output from the TPMs.

\subsection{Initial bootstrap}

Because it is necessary to both measure the unknown differential delays, and to confirm the signal path mappings, data from a dominant compact unpolarised radio source is required. For AAVS1, because the array is so compact, and because the antennas are sensitive to the entire sky, this is not always straightforward. The Sun is a known strong radio source, and when it is in a `quiet' mode, the flux density of the Sun is stable and well known.
The angular size of the radio Sun at frequencies of interest is approximately 1 degree, hence the quiet Sun is a compact strong unpolarised radio source for AAVS1 and can be used for initial bootstrapping.
The quiet Sun has well measured flux density over several decades of frequency \citep{2009LanB...4B..103B}, and we use this to set the flux scale for all of our data.
During all AAVS1 observations, the Sun was in solar minimum and showed no signs of abrupt amplitude variations, which would indicate an active state.

Using the channelised voltage capture mode of the local M\&C system, approximately 2\,s of voltages from each antenna were captured from a single coarse channel centred on 159.375\, MHz.
This frequency was used because it is known to be free from interference and is in the EoR key science range of interest.
The voltage data were correlated offline with a software correlator using 32 frequency channels so that frequency structure in both amplitude and phase is visible.
The data were then phased towards the Sun using the known (measured) locations of the antennas and the location of the Sun at the time of observation. The data were then exported in `uvfits' format and subsequently processed in the \textsc{miriad} data processing suite \citep{1995ASPC...77..433S}. Nine malfunctioning (see Sect. \ref{sec:testandverification}) or dead antennas were flagged at this stage.

Initial inspection of the data showed that the visibility phase had very little variation with frequency (typically less than 20 degrees across the coarse channel) on most baselines, which confirmed expectations that differential delays between antennas were small.
The next step was to apply a standard radio astronomy calibration to the data assuming a point source at the phase centre. Baselines shorter than five wavelengths were excluded.
The process revealed four antennas whose polarisations had been accidentally swapped during the station installation and this was corrected in software for subsequent processing by re-labelling the correlator outputs.

Having confirmed there are no large uncorrected delays, the real-time correlator was used to collect two hours of data centred on solar transit. Again, the data were phased to have the Sun at the phase centre.
Since the Sun is a compact source to AAVS1 and is at the phase centre, we expect the phase for all baselines to be constant in time, except for the shortest baselines, which are affected by large-scale structures in the sky.

\begin{figure*}
	\includegraphics[width=\textwidth]{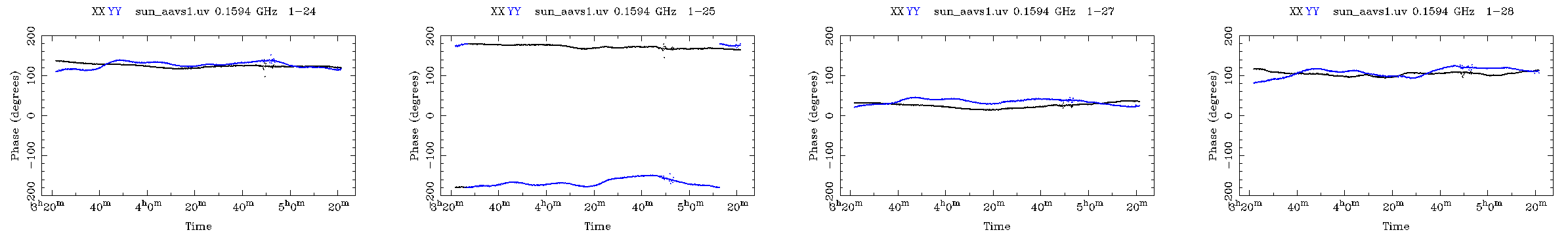}
    \caption{Example phases of AAVS1 uncalibrated visibilities for 2 hours of data taken around midday on 2018 April 5 at 159 MHz, with the Sun at phase centre.}
    \label{fig:example_phases}
\end{figure*}

Example uncalibrated visibility phases versus time for the two-hour dataset are shown in Fig. \ref{fig:example_phases} for some baselines. For all but the shortest baselines, the visibility phases are mostly flat as a function of time, consistent with a dominant compact source at the phase centre. Long baselines that have rapidly varying phase would be indicative of a signal path mapping error; however, no problems of this kind were seen for AAVS1 besides the polarisation swap problem mentioned above.
At this point, the signal mappings have been confirmed and the array functions correctly as an interferometer with no large uncorrected delays.

\begin{figure*}
	\includegraphics[width=\textwidth]{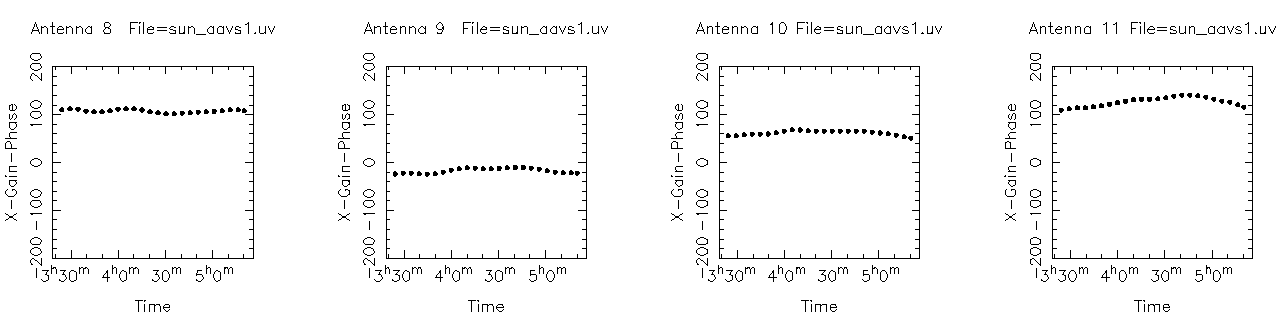}
    \caption{Example antenna-based phase calibration solution for 2 hours of data taken around midday at 159 MHz. An arbitrarily selected antenna near the eastern side of the array was chosen as the phase reference.}
    \label{fig:example_cal_phase}
\end{figure*}

The data were then calibrated assuming a compact source at the phase centre with a solution interval of 10 minutes. Example antenna-based calibration solutions are shown in Fig. \ref{fig:example_cal_phase}.
The calibration solutions show that the system is generally stable, but show antenna-based variations that are discussed in more detail below.

\begin{figure*}
\centering
	\includegraphics[width=0.32\textwidth]{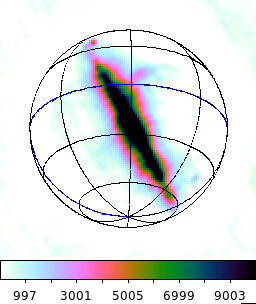}
	\includegraphics[width=0.32\textwidth]{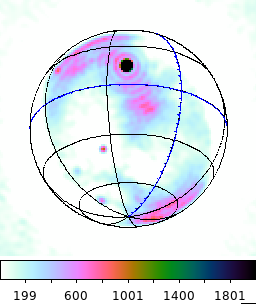}
	\includegraphics[width=0.32\textwidth]{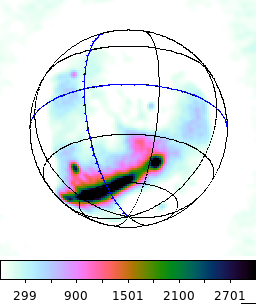}
	\includegraphics[width=0.32\textwidth]{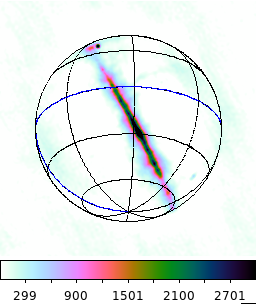}
	\includegraphics[width=0.32\textwidth]{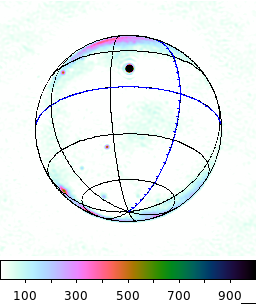}
	\includegraphics[width=0.32\textwidth]{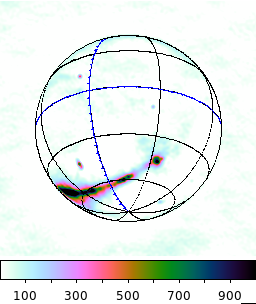}
	\includegraphics[width=0.32\textwidth]{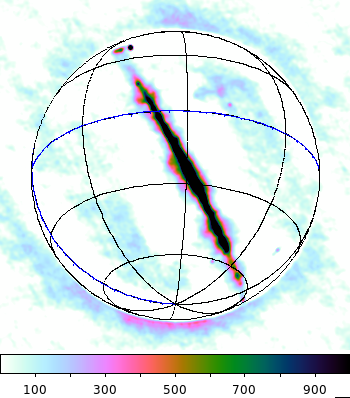}
	\includegraphics[width=0.32\textwidth]{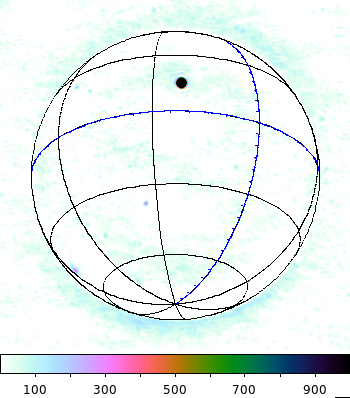}
	\includegraphics[width=0.32\textwidth]{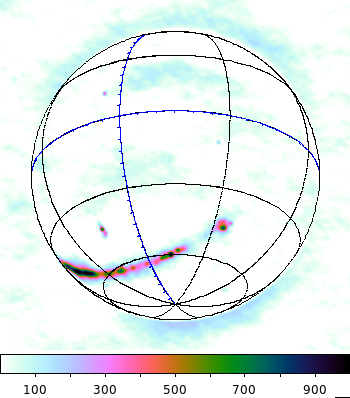}
    \caption{Example sky images made from 2.3\,s snapshots of data taken on 2018 April 26. Rows show frequencies 105, 200 and 308\,MHz (from top to bottom). Columns show LSTs approximately 18, 2 and 10 hours (from left to right). The bright source in the central column images is the Sun, which was used as a phase and flux density calibrator. The flux scale is Jy bm$^{-1}$ normalised at zenith, but the images have not been primary-beam corrected, hence they show beam-weighted apparent flux density of the sky in the snapshots, i.e. what the array saw.
    Note the colourbars are not all the same, and the scale has been set to highlight weaker sources and the general very high dynamic range of the snapshots, while stronger sources are saturated.}
    \label{fig:example_sky_image}
\end{figure*}

In an additional test of system stability and system mappings, the array was used for interferometric imaging of the full sky using conventional phase transfer techniques. Snapshot datasets of duration 2.3\,s were taken over a 24\,h period at several frequencies on 2018-04-26, and the dataset closest to noon was used for phase calibration for each frequency. Again, the Sun was used as a strong compact phase and flux density calibration source and baselines shorter than $4\lambda$ (for 105\,MHz) or $5\lambda$ (for 200 and 308\.MHz) where discarded. The calibration solutions were applied to the full time window, and each snapshot was independently imaged over the full sky with the zenith as phase centre.
Example images are shown in Fig. \ref{fig:example_sky_image} for the XX polarisation.
The lack of imaging artifacts around bright sources in the local sidereal time (LST) 10\,h and 18\,h images also demonstrates that the array is phase stable with a single phase calibration solution.
To generate the images, a modest amount of cleaning (5000 iterations) was performed on each snapshot. The images correspond well with expectations from all-sky maps with both large-scale emission from the galaxy and brighter compact sources clearly visible in the correct location. Low-level image artefacts from missing very short baselines are apparent in the LST 18\,h, 308\,MHz image as well some low-level imaging artefacts near the horizon near in the LST 2\,h, 105\,MHz image.

\subsection{Residual delay calibration}
\label{sec:resid_delay_cal}
\begin{figure*}
	\includegraphics[width=\textwidth]{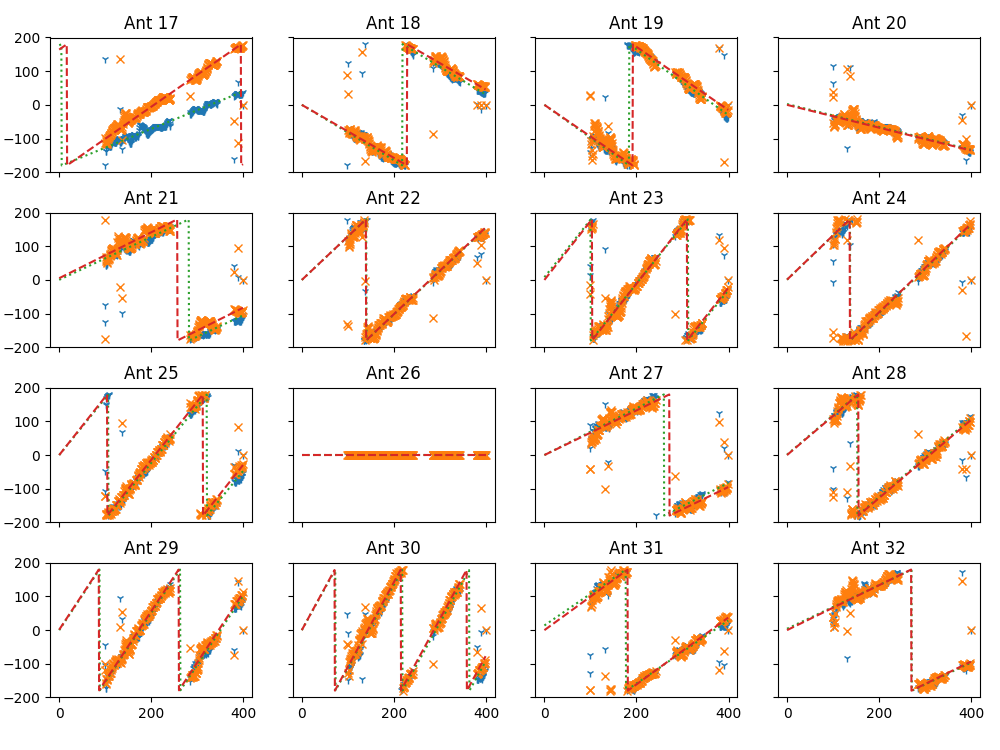}
    \caption{Example antenna-based phase calibration solution over a broad range of frequencies. Antenna 26 was offline. Data below 100\,MHz and in frequency bands known to be affected by persistent RFI have been excluded. Individual points are shown for the two polarisations (orange `X' and blue `Y') as well as the predicted phase from a fitted delay for each polarisation, shown by the dot or dashed red and green lines.}
    \label{fig:example_cal_delay}
\end{figure*}

Given that the Sun can be used as a dominant compact source for calibration, the same process was used across the full frequency range, where antenna-based calibration solutions were generated independently for each coarse channel. Example phase solutions are shown in Fig. \ref{fig:example_cal_delay}.
Data below 100\,MHz (where the Sun is not dominant) and in frequency bands known to be affected by RFI have been excluded.

It is clear that the antenna-based phase solutions follow a linear function of frequency. This behaviour is consistent with the antenna calibration phase being dominated by residual uncorrected delays (presumed to be differential cable lengths over the full signal path between antenna and TPM) and hence the antenna-based calibration phase can be described by a single delay for each polarisation. It is also clear that for most antennas, the delay between the two polarisations is very similar, which is consistent with the two polarisations sharing the same optical fibre.

Antenna 17 in Fig. \ref{fig:example_cal_delay} also highlights two additional issues. Firstly, not all antennas have identical delays between the two polarisations (also apparent for Antenna 21) and secondly, Antenna 17 has phase $180^{\circ}$ at zero frequency.
Three other antennas also exhibited the property of having $180^{\circ}$ phase at zero frequency.
Follow-up inspection of these antennas showed that they had accidentally been installed with their polarity reversed, either by rotation of the trumpet on the antenna or rotation of the entire antenna assembly.

Based on these observations, a phase calibration model for each antenna was formed with two parameters for each polarisation: a differential delay and a 0\,Hz intercept, which is constrained (but not forced) to be either 0 or $180^{\circ}$. The phase on a polarisation $p$ for antenna $i$ is given by $\phi_{i,p} = 2 \pi \nu t_{i,p} +\phi_{i,p,0}$ where $\nu$ is the observing frequency, $t_{i,p}$ is the fitted delay and $\phi_{i,p,0}$ is the fitted 0\,Hz intercept.
The phases predicted by the fitted models are also shown in Fig. \ref{fig:example_cal_delay} as the dotted and dashed lines. The fitted phases show excellent agreement with the data and the small deviations of the data from the fitted model are discussed in more detail below.

\subsection{Calibration accuracy}
\label{sec:cal_accuracy}
Ideally, with a dominant compact source at the phase centre, both the uncalibrated visibility phases and gain solutions should be constant with time, hence the residual phase variations seen in the raw visibilities and antenna-based gain solutions (Figs. \ref{fig:example_phases} and \ref{fig:example_cal_phase}) were seen as an issue, considering that the SKA-Low requirement at the time called for better than one degree accuracy in the beam-forming weights.

As well as considering physical effects such as temperature variations in fibres, LNAs, and FEMs, mutual electromagnetic coupling between antennas was seen as potentially causing the observed variations. Mutual coupling is a well known effect in aperture arrays with regular or closely spaced antennas \citep{Mailloux3rd,Warnicketal}.

The data presented here cannot unambiguously demonstrate that mutual coupling was occurring; however, concern about the presence and effects of mutual coupling has spawned a significant amount of work in the wake of AAVS1. This work includes details of the modelling itself \citep{8879294,2020Steiner}, estimating the magnitude of the effects of mutual coupling \citep{8879294,8739902,9232307,2020MNRAS.496..933B}, verifying the simulations \citep{9232190}, and estimating the impact of the coupling on calibration and downstream signal processing \citep{9232295,2021MNRAS.505.1485J}.

The station beam that is formed by application of biased calibration solutions will suffer some degradation from the ideal. However phase errors of the order of magnitude 5--10$^{\circ}$ seen in Figs. \ref{fig:example_phases} and \ref{fig:example_cal_phase} will not significantly deform the actual station beam shape, which is confirmed in Sect. \ref{sec_beamforming}, and will have a negligible effect on sensitivity \citep{2017PASA...34...34W}.


For the calibration results presented in this paper, the delays computed for antennas in Sect. \ref{sec:resid_delay_cal} are not biased because the delay solution is fit over a broad range of frequencies, hence the bias averages out (clearly seen in Fig \ref{fig:example_cal_delay}). A similar conclusion was reached by \citet{9232295}, although we note that in this case the delay solutions were generated from the most ideal data, based on the Sun around solar transit.

\subsection{Calibration long-term stability}

The procedure to solve for antenna-based delays was run daily around midday for several months to assess the stability of the delay solutions over time.

\begin{figure}
	\includegraphics[width=\columnwidth]{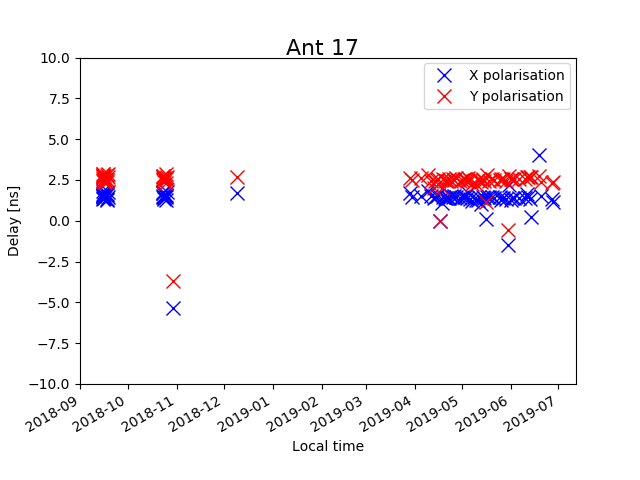}
    \caption{Example delay solutions for antenna 17 as a function of time in the interval from August 2018 to July 2019. This particular antenna has slightly different delays (i.e. cable length) for X and Y polarisations, which, except for a few outlier points, remained very stable over time.}
    \label{fig:example_delaysol}
\end{figure}

\begin{figure}
   	\includegraphics[width=\columnwidth]{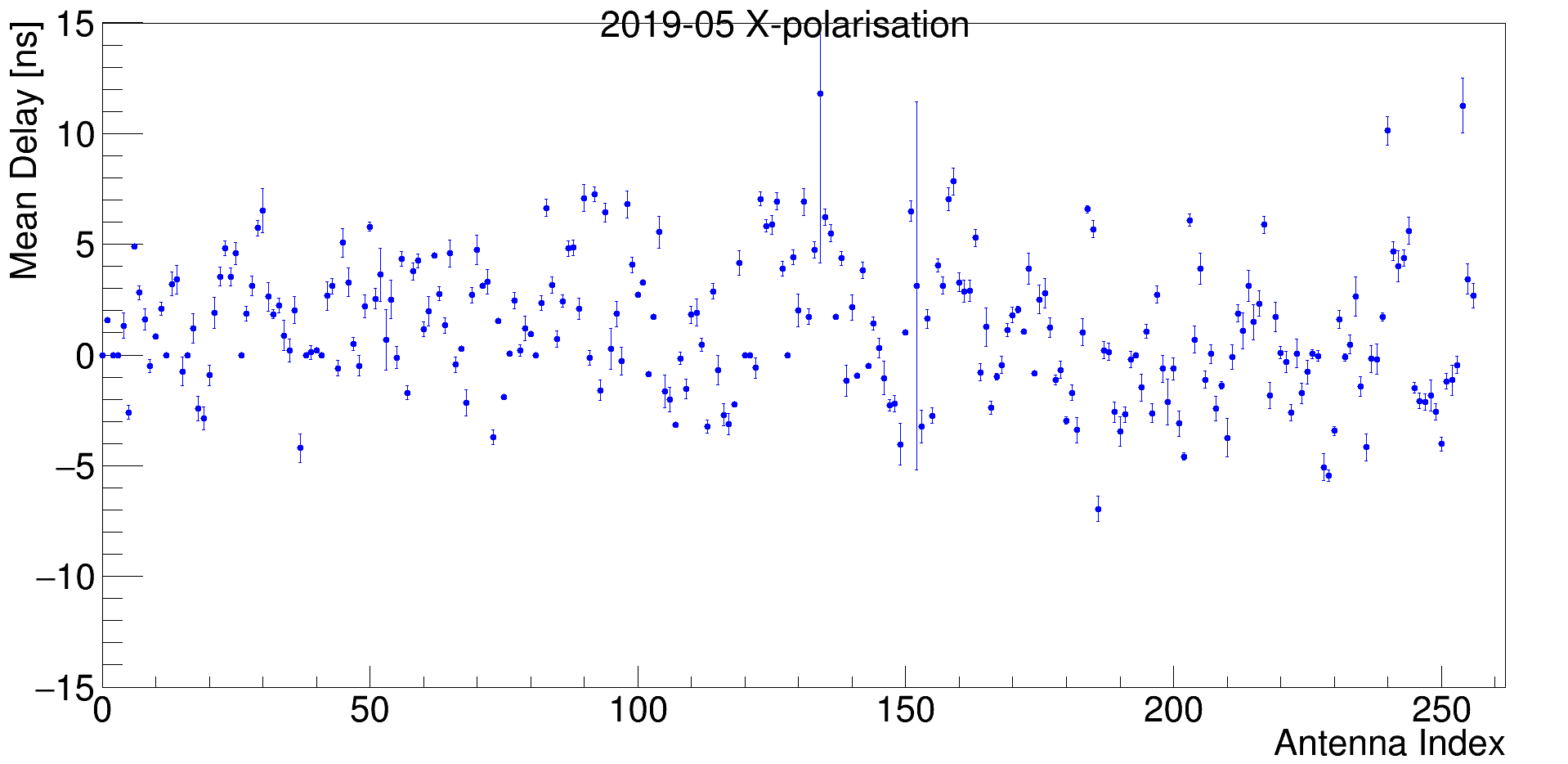}
   	\includegraphics[width=\columnwidth]{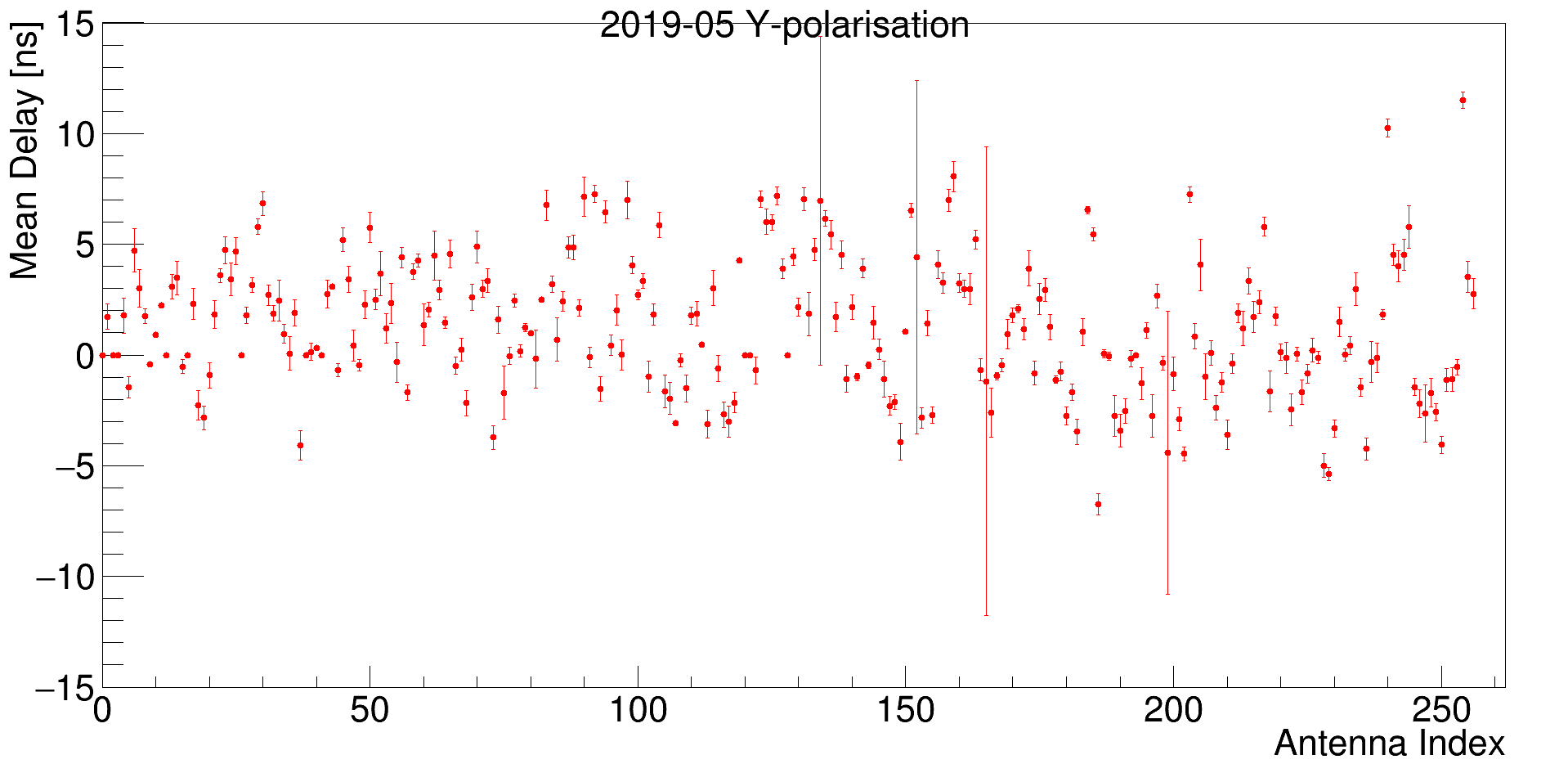}
    \caption{Mean and standard deviation (error bar) of delays for each antenna in May 2019. The upper image (blue points) is X polarisation and the lower image (red points) is Y polarisation.}
    \label{fig:delaysol_all}
\end{figure}

Figure \ref{fig:example_delaysol} shows example delay solutions for an antenna for data taken over approximately 9 months between 2018 and 2019. The data show that the delay solutions are very consistent with time. The standard deviation of the delay solutions over all antennas is shown in Fig. \ref{fig:delaysol_all}, where the obvious outliers have been excluded.
This figure shows that the bulk of the antennas have small deviation to their fitted delay values, of order 1\,ns or less over the nine-month timescale. 

\section{Beamforming}
\label{sec_beamforming}

Real-time station beam-forming is implemented in the 16 TPMs as a daisy chain, coherently summing the voltages from individual antennas within a single TPM to form a so-called tile beam then passing the result to the next TPM as discussed in Sect. \ref{sec:digital_path}.
The resulting complex voltages of the tile beam are passed from the first TPM in the chain to the second TPM which forms its own tile beam, adds the two beams together, passes to the third TPM  and so on until the last TPM in the chain sends the resulting station beam (complex voltages) to data acquisition computer. Details of the station signal processing can be found in \cite{2017JAI.....641015C}.


For AAVS1, calibration coefficients were applied as phase corrections to each coarse channel, which formed a coherent station beam. By default the station beam points towards the zenith, but can be steered by applying additional phase corrections for each antenna for each coarse channel, depending on the look direction.

In order to verify the real-time beam-forming procedure data, were collected simultaneously in station beam and channelised voltage modes. An offline beam-forming procedure was used to beam-form the data collected in the channelised voltage mode in order to compare total power collected over 24\,hours (typically at zenith) in both modes. This allowed us to debug the real-time beam-forming procedure and converge on perfect agreement between 24\,hour total power curves versus time (`light curves') from the real-time and offline beam-forming procedure.

Both light curves were compared against simulations. At a given time $t$ the modelled antenna temperature ($T_{ant}$), which is the power received by the station beam, can be calculated as:
\begin{equation}
T_{ant}^{model}(t,\nu) = \frac{\int_{4\pi} B(\nu,\theta,\phi) T(t, \nu,\theta,\phi) d\Omega}{\int_{4\pi} B(\nu,\theta,\phi) d\Omega}, 
\label{eq:sky_integration}
\end{equation}
where $B(\nu,\theta,\phi)$ is the AAVS1 station beam power pattern, $T(\nu,\theta,\phi)$ is the sky brightness temperature from the sky model at frequency $\nu$ and pointing direction $(\theta,\phi)$.

In our simulations we used two sky models. The so-called HASLAM map at 408\,MHz \citep{1982A&AS...47....1H} was scaled down to observing frequencies using spectral index $-2.5$, or the global sky model \citep[GSM; ][]{2008MNRAS.388..247D} which provides sky maps at the requested frequency.
The system temperature was calculated as $T_{sys}(t,\nu) = T_{ant}^{model}(t,\nu) + T_{rcv}(\nu)$, where receiver temperature, $T_{rcv}(\nu)$, was obtained from laboratory measurements
and was assumed to be constant in time.

\begin{figure*}
\begin{center}

\includegraphics[width=1.0\textwidth]{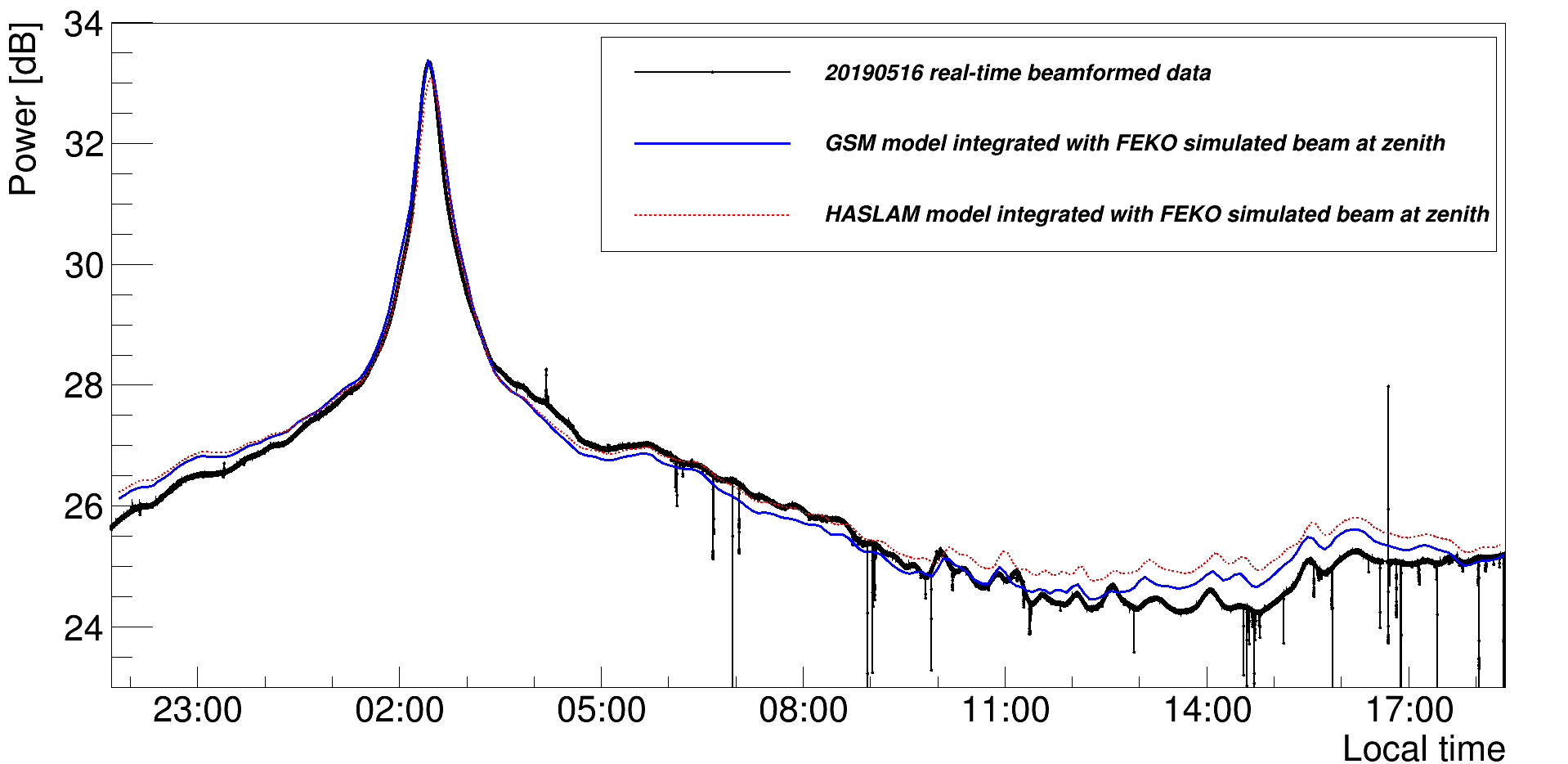}


 \includegraphics[width=1.0\textwidth]{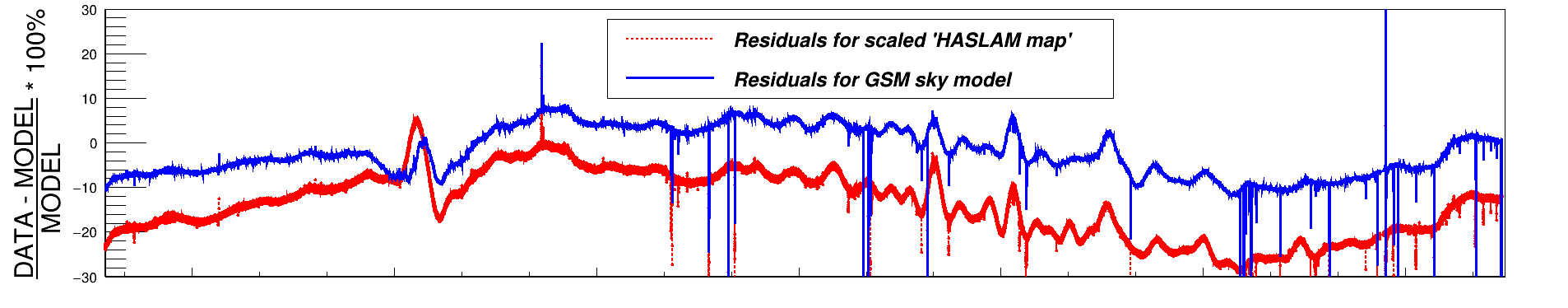}
\caption{Total power in decibel-kelvin scale (calculated as 10\,$log_{10}(T_{sys})$) of the AAVS1 station beam at zenith at 159.375\,MHz, as a function of local time (top image). The black curve is the data points from the AAVS1 real-time station beam. The solid blue curve is the model prediction using the GSM and the dashed red curve is the prediction using the HASLAM map scaled to our observing frequencies using a spectral index of $-2.5$. Both predictions incorporate a model of the AAVS1 station beam pointed at zenith. The lower plot shows the difference between the data and simulated curve using GSM sky model (maximum difference $\approx$15\%) and scaled HASLAM map (maximum difference $\approx$20\%). The vertical spikes are due to RFI or lost packets, which used to occasionally happen when there were other processes running on the AAVS1 server during the data acquisition.}
\label{fig_beamform_data_vs_simul}
\end{center}
\end{figure*}

In order to compare the data to the simulated light curves, the data were normalised to simulations at the maximum of the light curve (approximately at the time of the Galactic centre transit).
Hence, the gain factor $g(\nu)$ was calculated as:
\begin{equation}
g(\nu) = \frac{ \biggl(T_{ant}^{model}(t_{peak},\nu) + T_{rcv}(\nu) \biggl) }{ P(t_{peak},\nu) }
\label{eq:power_vs_tmodel}
\end{equation}
where $P(t_{peak},\nu)$ is the station total power at the approximate time of the Galactic centre transit ($t_{peak}$) and all the data points were multiplied by this gain factor.
We assumed that $g(\nu)$ did not change over the time of the observations (typically 24\,hours). The comparison of data collected on the night 2019-05-16/17 and simulation for the time interval of about 20 hours is shown in Figure~\ref{fig_beamform_data_vs_simul}. 
This particular figure shows the data beam-formed in real time, but it was verified that the real-time beam-formed light curve was exactly the same as the one formed offline, therefore only one of them is shown.

The data matches simulation predictions to within 15\% in the case of GSM model and 20\% in the case of the scaled HASLAM map, which reflects the fact that the GSM model should be a better representation of the true sky at these frequencies.
The agreement between the predicted and measured light curves confirms that the beam-forming is working correctly. In particular, the height and width of the Galactic centre peak, and the smaller bumps and wiggles in the curves line up very well. 
Given that an idealised model of both the sky and the station beam were used for the predicted light curve, the small differences between the predicted and measured light curves are expected.


\section{On-sky sensitivity}\label{sensitivity}

To investigate the sensitivity that can be achieved with AAVS1 in the two linear polarisations $X$ and $Y$, we used solar observations from 2018 April 26 when the Sun was in an inactive phase. Data were obtained at $17$ separate frequencies in voltage capture mode, spanning the range $58.6$--$308.6$\,MHz. Each frequency had a bandwidth of $0.926$ MHz. The integration time was $2.26$\,s, comprising two consecutive $1.13$\,s time steps.
The data were captured close to midday (Sun elevation $49\fdg9$).
All of the observing frequencies were cycled through over a $15.5$\,min period, starting at the lowest frequency first.
To obtain interferometric visibilities, the voltage-capture data were then correlated offline into $32$ channels per observing frequency, and the phase centre was shifted to the position of the Sun. A total of ten antennas in the station were offline during the observations; these were the known bad antennas previously mentioned in Sect.~\ref{sec_calibration}.   

\subsection{Calibration of solar data}

Gain and bandpass calibration was carried out using {\sc miriad}. We used a model for the quiet Sun based on the tabulated Stokes-$I$ flux densities in \cite{2009LanB...4B..103B}; between $50$ and $350$ MHz, the solar flux density varies from $5.4$ to approximately $180$ kJy.
To determine the solar flux density at a particular AAVS1 observing frequency, we interpolated between the various frequency pairs in \cite{2009LanB...4B..103B} using the corresponding two-point spectral index. We used a short baseline cutoff for the calibration ($3$, $4$, or $5\lambda$, depending on the specific observing frequency) due to the significant flux density on shorter baselines from the Galaxy, and also sometimes from RFI from the direction of the horizon. It should be noted that the quiet Sun is unpolarised \citep{2009LanB...4B..103B}, and this was taken into account when calibrating the data.  \citet{2009LanB...4B..103B} state that the overall uncertainty on their flux scale is 10\%.

The data at $58.6$ and $74.2$ MHz could not be successfully calibrated using the Sun owing to the sky brightness temperature at these frequencies, as well as problems associated with the horizon RFI. Moreover, the data at $246.1$ and $261.7$\,MHz were generally affected by RFI. This left $13$ frequencies (range $89.8$--$308.6$\,MHz) for which we could measure the on-sky sensitivity of AAVS1.

\subsection{Difference imaging}\label{section:difference imaging}

Because the AAVS1 fringe rate is sufficiently slow, barring variability from, for example, an astrophysical transient or source of RFI, the sky brightness distribution will not change significantly between the two consecutive $1.13$\,s time steps (although see the discussion later in this section). Therefore, to obtain a sensitivity measurement at each observing frequency, we made an image for each time step, and then subtracted the first from the second to create a difference image. Generally speaking, this simple approach is useful because, in the ideal case, it will remove the astrophysical sources, calibration artefacts and any background levels or ripples, as well as circumventing the limitations in sensitivity owing to classical and/or sidelobe confusion. An RMS can then be measured from the noise-like difference map. For two input images with underlying Gaussian noise $\sigma$ Jy beam$^{-1}$, the corresponding difference image will have an RMS of $\sqrt{2}\sigma$\,Jy beam$^{-1}$. 

Imaging was also carried out in {\sc miriad}; we used natural weighting to optimise the sensitivity. We found that light {\sc clean}ing the single-time-step images was necessary before differencing: tiny variations in the dirty beam patterns between the two time steps were significant because the Sun is a very strong source at our observing frequencies, and without deconvolution, the difference-image sensitivity would have been biased upwards by residual sidelobe structure. In a number of cases, $500$ {\sc clean} iterations in a small box surrounding the position of the Sun was sufficient. Otherwise, we used an unconstrained {\sc clean} across the full image with $5000$ iterations to remove as many of the sidelobes as possible.   

\begin{figure*}
\begin{center}
\includegraphics[width=0.32\textwidth]{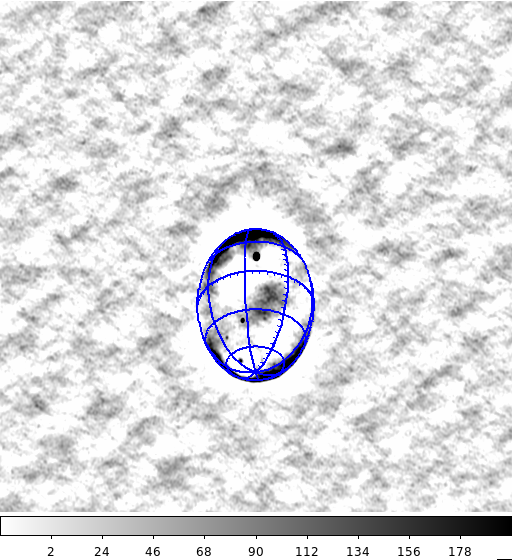}
\includegraphics[width=0.32\textwidth]{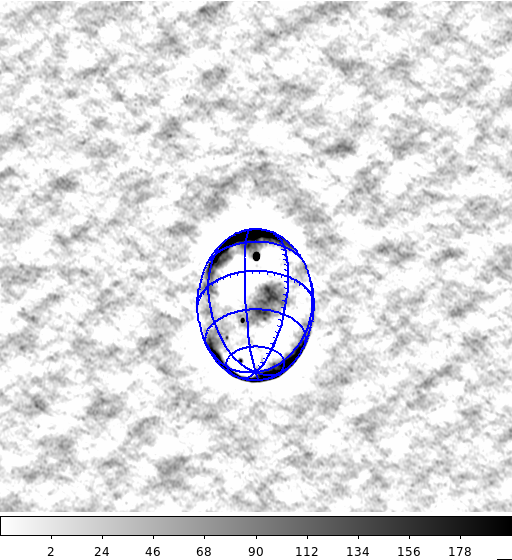}
\includegraphics[width=0.32\textwidth]{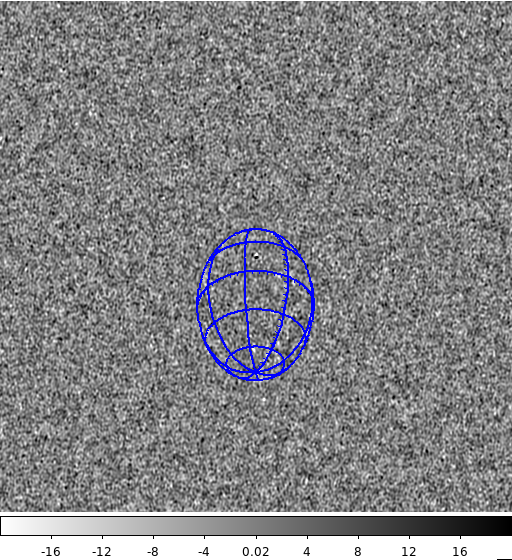}
\caption{Maps for linear polarisation $X$ at $199.2$\,MHz from the solar observation on 2018 April 26 (Sect.~\ref{sensitivity}), shown in the native slant orthographic projection of the snapshots. Time step one is the left panel, and the middle panel is time step two. Subtracting the left panel from the middle panel gives the difference map shown in the right panel (we note the different colour bar scales; the colour bar has been deliberately set to show low-level signal vs. the $\sim50000$\,Jy Sun).
The Sun is the point source at the centre of both the left and middle panels; after difference imaging, positive and negative residuals remain at the position of the Sun, but only at the $0.05$ per cent level in this particular case. {\sc clean}ing was restricted to a small box surrounding the Sun (see Sect.~\ref{section:difference imaging}). The difference image is clearly noise-like, illustrating the effectiveness of this technique.}
\label{fig:difference_map}
\end{center}
\end{figure*}

Separate sets of images were made for the $X$ and $Y$ polarisations; a summary is presented in Table~\ref{table:sensitivity}. The angular resolution in right ascension varied from about $5\fdg2$ to $1\fdg5$ full width half maximum over the frequency range. The resolution was about $30$ per cent coarser in declination. Example images are shown in Fig.~\ref{fig:difference_map}. The difference images often had artefacts at the position of the Sun owing to imperfect subtraction; typically, these were significantly less than one per cent of the solar flux density, however.
Additional subtraction residuals, sometimes related to the horizon emission or other instances of terrestrial and satellite-based RFI, were also seen at some observing frequencies. The most prominent example was one source of RFI varying by nearly 1\,kJy in apparent flux density between the two $1.13$\,s time steps at $136.7$ MHz in $X$, which is a band known to be affected by satellite-based RFI. Generally speaking, however, the majority of the difference images had large, noise-like regions (e.g. Fig.~\ref{fig:difference_map}). We note that the emission near the northern and southern horizon in Figure~\ref{fig:difference_map} was seen at several frequencies and is due to a combination of missing zero spacing information and strong horizon response of the SKALA2 antenna at some frequencies. It is clear from Figure~\ref{fig:difference_map} that this emission subtracts in the difference image like the rest of the sky.

We imaged significantly beyond the physical horizon of the telescope to take advantage of the fact that interferometric images have noise equally distributed across the entire image before primary beam corrections are made. The regions of the image beyond the horizon contain only noise and sidelobes.
Because our maps do not have standard primary beam corrections applied to them, we could then measure the average RMS in the outer part of each difference map where any artefacts from the subtraction process were significantly weaker, and the difference maps more noise-like. The uncertainty per RMS estimate was mostly $\lesssim$ $5$ per cent, although for the more problematic difference images, it was about $10$--$15$ per cent. Combined with the 10 per cent uncertainty from the solar flux scale, we conservatively estimate the overall uncertainty to be 15 per cent.

After calculating the RMS noise levels, we then made corrections (as determined from simulations) to account for the position of the Sun in the antenna element beam for a given linear polarisation, frequency and time of observation. This step ensured that the calibrated Sun had the correct apparent, rather than absolute flux density. In the majority of cases, an additional correction was needed to normalise the gain to zenith as per the SKA specification (although usually the percentage difference was small). The net multiplicative factors ranged from $0.257$--$1.305$; multiplying the image noise level by the appropriate factor in each case scaled the corresponding difference-image sensitivity to a zenith value, for comparison with SKA-Low specifications. 

Our corrected RMS noise levels from the difference images are presented in Table~\ref{table:sensitivity}. We also converted each noise level to a measured station system equivalent flux density (SEFD) using the following equation:
\begin{equation}
\label{eqn:sensitivity_SEFD}
{\rm SEFD_{measured}} = {\rm RMS_{difference\:image}} \times \sqrt{\frac{\Delta \nu \Delta t}{2}} \times \frac{246}{256}.  
\end{equation}
In Eq.~\ref{eqn:sensitivity_SEFD}, the bandwidth per frequency $\Delta \nu = 0.926$\,MHz, and the integration time per time step $\Delta t = 1.13$\,s. As discussed above, the factor $\sqrt{2}$ results from the assumption that the two input images used to make a particular difference map have identical Gaussian noise characteristics. Furthermore, the fraction $246/256$ is a first-order correction for the fact that ten antennas were offline during the observations.      
\subsection{Comparison of measured zenith sensitivities with SKA-Low requirements}\label{difference imaging comparison requirements}

In Table~\ref{table:sensitivity}, we have also calculated the target SEFD, ${\rm SEFD_{requirement}}$, at each of our observing frequencies using the current SKA-Low requirements \citep[specifically SKA1-SYS\_REQ-2135;][]{SKA1L1req}.
To do this, we interpolated between the $A/T$ requirement values for the full SKA-Low (i.e. $512$ stations) at the given frequencies in \citet{SKA1L1req}, using a `not-a-knot' cubic spline function, which is specified in SKA1-SYS\_REQ-2135 as the function to be used for interpolation across the SKA-Low frequency range. We then divided by $512$ to obtain values for a single station. To calculate ${\rm SEFD_{requirement}}$, we used the standard equation
\begin{equation}
{\rm SEFD_{requirement}} = \frac{2k}{(A/T)_{\rm station}},
\end{equation}
where $k$ is the Boltzmann constant. 

For the 2018 April 26 observation, the median ratio ${\rm SEFD_{measured}}/{\rm SEFD_{requirement}}$ was $1.5$ in both $X$ and $Y$. Although the sensitivity requirement was not formally met at any of the observing frequencies, it is, for example, encouraging that the discrepancy was $< 30$ per cent at $89.8$ and $199.2$\,MHz in $X$, as well as at $168.0$, $199.2$, $293.0$ and $308.6$\,MHz in $Y$. The change in ${\rm SEFD_{measured}}$ as a function of frequency did not agree one-to-one with the expected sensitivities for the SKALA2 antenna \citep[e.g.][]{2017MNRAS.469.2662D}, although there were some hints of consistency for particular frequency ranges. While it was clear that some difference images were poorer in quality than others, unfortunately other suitable AAVS1 datasets with short integration times were not available to conduct similar tests to much better establish the change in measured sensitivity across the SKA-Low observing frequency range. As part of ongoing commissioning for the current generation of SKA-Low prototype stations, voltage-capture observations are being taken regularly to thoroughly assess the sensitivity capabilities, and with even finer time steps (e.g. $0.14$ s) to reduce the effects of sky rotation. 

\begin{table*}
\begin{minipage}{1.0\textwidth}
\centering
\caption{AAVS1 zenith-sensitivity measurements for the $X$ and $Y$ linear polarisations from solar observations carried out on 2018 April 26. SKA-Low requirements have been taken from \citet{SKA1L1req} and converted into values for a single station. See Sect.~\ref{sensitivity} for further details.}
\begin{tabular}{cccccc}
\hline
Frequency & \multicolumn{2}{c}{$\rm{RMS_{difference\:image}}$} & \multicolumn{2}{c}{$\rm SEFD_{measured}$} & SKA $\rm SEFD_{requirement}$  \\
& $X$ & $Y$ & $X$ & $Y$ & per linear polarisation per station  \\
(MHz) & (Jy beam$^{-1}$) & (Jy beam$^{-1}$) & (kJy) & (kJy) & (kJy) \\  
\hline
$89.8$ & $7.7$ & $18.3$ & $5.4$ & $13$ & $4.17$ \\
$105.5$ & $5.5$ & $13.9$ & $3.8$ & $9.7$ & $2.84$ \\
$121.1$ & $7.3$ & $8.5$ & $5.1$ & $5.9$ & $2.46$ \\
$136.7$ & $8.9$ & $9.8$ & $6.2$ & $6.8$ & $2.41$ \\
$152.3$ & $5.1$ & $5.6$ & $3.5$ & $3.9$ & $2.35$ \\
$168.0$ & $5.2$ & $3.9$ & $3.6$ & $2.7$ & $2.29$ \\
$183.6$ & $7.1$ & $5.0$ & $4.9$ & $3.5$ & $2.27$ \\
$199.2$ & $4.1$ & $3.8$ & $2.8$ & $2.6$ & $2.27$ \\
$214.8$ & $6.3$ & $5.0$ & $4.4$ & $3.5$ & $2.29$ \\
$230.5$ & $4.4$ & $4.8$ & $3.1$ & $3.3$ & $2.33$ \\
$277.3$ & $7.3$ & $4.7$ & $5.1$ & $3.3$ & $2.45$ \\
$293.0$ & $5.2$ & $4.5$ & $3.6$ & $3.1$ & $2.49$ \\
$308.6$ & $5.3$ & $4.6$ & $3.7$ & $3.2$ & $2.54$ \\
\hline
\end{tabular}
\label{table:sensitivity}
\end{minipage}
\end{table*}

\section{Lessons learned}
\subsection{Procurement and construction}
The deployment of the LFAA stations will be a significant factor in the cost and efficacy of the SKA-Low construction phase. Assembly and installation of the currently planned 131,072 individual antennas -- and the supporting and connecting elements that combine with the antennas to make up a station -- will be a complex, labour intensive exercise. A deployment of this (numerical) scale is unprecedented in radio-astronomy, and has few close analogues from which lessons can be directly drawn. These factors saw a high level of perceived risk attached to the LFAA deployment early in the SKA design phase. 

AAVS0 and AAVS\,0.5 did not trigger the challenges of scale that characterise SKA-Low. Though still of a modest size compared to SKA1, at 256 antennas (400 were produced) AAVS1 forced an important transition from small-scale, institute-based production and quality control to a regime that is more applicable to SKA. The kinds of production, transport, assembly and installation procedures and techniques that will be required for SKA were exercised for the first time.

The deployment, installation and commissioning of AAVS1 clearly bore out the significant drivers of the LFAA deployment cost and schedule; and identified a variety of issues arising from preparatory activities including procurement, manufacture and transport. A representative sample of these issues are noted in the following paragraphs.

\subsubsection{System thinking in appreciating cost.}
In the early part of the LFAA design activity there was a propensity for costing exercises to focus on a component-level build up to a system-level (LFAA) bill-of-materials. A range of factors contributed to this approach, including the level of decomposition of the design and corollary assignment of engineering responsibility; and the highly distributed nature of the design team. Chief among the many risks and issues inherent in this approach to costing are the inability to (easily) optimise cost across design elements at the system-level; an increased likelihood of technical and cost gaps; and the tendency to overlook effort-based cost inputs such as deployment, assembly and installation. Planning for AAVS1 forced the design team to break out of this sub-optimal mode of thinking.  

\subsubsection{Quality control.}
In AAVS1 the (analogue RF on optical fibre) signal path between antennas in the field and the signal processing system (SPS) installed in the MRO control building is provided by two 576-core optical fibre cables (approximately 5\,km in length). Like any commodity item, a variety of optical fibre products are available in the market place that, theoretically, conform to standards-based specifications. These products span a considerable price range that, given the quantity of fibre required at SKA-Low scale, translates to fibre accounting for between 5-15 per cent of the capital budget (design phase target) for LFAA. The need for two cables--reflecting a total fibre count driven by a requirement for two fibres per antenna prior to the adoption of optical components utilising WDM, and a total antenna count of up to 400--afforded the opportunity to evaluate more than one source of supply with a view to understanding the price variation in the market.

An agent was appointed to procure both cables, to the same specification, on behalf of the AAVS team. One of the cables was procured from a premium vendor whose offering was towards the top end of the price range returned by a market survey. The other was procured from a vendor whose product was towards the lower end of the price range. The agent's role included verification that the cables supplied were compliant with the specification issued. Both cables were inspected prior to departing their country of manufacture. The cheaper cable was found to be non-compliant with a number of important, non-performance-related, requirements. The agent ensured that all non-conformances were rectified -- which required considerable re-work of the cable cladding -- before accepting it and allowing it to be embarked for shipment to Australia.

This is as an excellent example of the false economy of accepting the lowest tendered price, and a reminder of the importance of due diligence and quality control in procurement. These measures can, cumulatively, represent a considerable investment and in a budget constrained environment they are a tempting target for savings. The AAVS1 experience stands as a positive demonstration of their value. 

\subsubsection{Installation test regime.}
Thorough component-level characterisation was a central tenet of the AAVS1 delivery programme. Collection of test data and results was a focus of the procurement, deployment and installation activities, to support AAVS1's mission as an engineering test system. A key example in deployment was the detailed characterisation of the optical-fibre (trunk) link between the AAVS1 station in the field and the MRO control building. This activity exposes the challenge of scale that is an important factor inherent in many SKA cost-schedule-quality tradeoffs.    

The 576-core optical fibre cables procured for AAVS1 were delivered with factory test certificates supported by test results. However, the handling and processes involved in transport and installation of the cables carry a level of risk of damage that demand that the fibre be retested following installation. For AAVS1, post-installation testing consisted of optical time‐domain reflectometry (OTDR) of every individual fibre core to confirm link connectivity, and to obtain an optical path length measurement. Each core was tested at three wavelengths--1310nm, 1550nm and 1625nm; with each measurement taking 30 seconds.
The resulting battery of 1728 individual tests (per cable) represent 15-hours' work for a team of two (30 person hours). Including a productivity factor to reflect realistic work rates, it is reasonable to budget two full work days, for two people, for each cable (40-hours of person effort per cable). The AAVS1 experience also showed that as much time again was spent moving the test setup between test-sites; confirming the correct fibre to test; cleaning and preparing fibre connectors for the test; and configuring and transferring files from the OTDR machine’s internal storage (which filled many times over during the testing). 

Extrapolating to SKA-Low -- where the equivalent fibre cables carry only half of the number of fibre cores as the AAVS1 cable -- more than 20,000 person hours of effort would be required to implement the same test regime. The (direct) cost of this liability, in the context of the SKA1 construction budget, could vary considerably depending on how the fibre test activity is procured and delivered.
While the cost will not be insignificant, it is not the only consideration associated with the fibre test regime. Arguably, the greater impact comes in the way the requirements of the testing -- including preconditions, access, duration and dependences -- impact the sequence and progress of SKA-Low construction activities. The fibre cables must have been installed in their final arrangements. The end points of each cable--one in the field and the other in a processing facility--must be accessible to the test team(s). Subsequent, system-level, integration and commissioning cannot proceed until the testing is completed, including the rectification of issues and confirmatory re-tests. These elements can have a significant influence on the overall construction schedule and cost.              

This example is not presented as an argument against thorough component level testing and characterisation. Rather, it is intended to highlight how, at scale, apparently minor tasks quickly accumulate to significant activities. When dealing at the (numerical) scale of SKA-Low, all installation operations, no matter how minor, need to be properly motivated. The utility of any test regime has to be weighed against the direct and consequential impacts it has on the programme.

\subsection{Practical issues, testing and verification}
\label{sec:testandverification}
In addition to the debugging that is always required following the first at-scale deployment of a prototype system, a number of issues were encountered that highlight significant practical considerations for the transport, deployment and in-field verification and maintenance of the LFAA.

\subsubsection{Auto-oscillation.} The initial full signal chain integrated testing was performed at the MRO, during the first installation site trip, in late 2016. Although initial test results with a few antennas met expectations, during the installation and commissioning of the second batch of antennas, in March 2017, several of the installed antennas started to show an auto-oscillation around 69\,MHz. 
The issue first had to be reproduced in the lab, then a list of possible remedies had to be identified, before further deployment and commissioning of AAVS1.
The issue was essentially due to the marginal stability (and high gain) of the LNA and RFoF hardware, combined with a positive feedback loop from the hybrid cable to the electronics in the antenna.
The project was delayed for 6 months as the issue was investigated and several measures were proposed to remedy it. These were implemented during a large (and long) site-trip in November 2017.

\subsubsection{Initial quality control of manufacturing}
Imperfect and non-homogeneous galvanisation of the antenna elements caused oxidation of the metal during transport and after a short time in the field, affecting performance and maintainability. This was later addressed by updating the manufacturing process.

\subsubsection{In field assembly}
The antenna, trumpet and hybrid cable system contains many small, fiddly components that are not well suited for in-field assembly or maintenance. In addition, some components that were assembled and unit tested prior to being shipped to the MRO developed problems (e.g. loose screws) during transport, requiring difficult in-field remediation. Compounding this problem was the issue that some components could easily be damaged by (for example) over-tightening screws or other means.

\subsubsection{Hybrid cable}
There were several issues with the hybrid cable and APIU concept. Handling 256 cables into a single box was very difficult, and risked damage to the fibres. The cables are relatively thick, and routing them between the randomly placed antennas was non-trivial. The excess cable lengths had to be carefully managed and coiled around antenna bases.

Reiterating lessons from the previous section, many of these are problems that only appear when components are deployed at scale and/or in the field, hence the importance of prototyping and iterative design updates.

\section{Future}
As the SKA1 pre-construction Phase has been wrapped up within the element (CDRs), most of the residuals of the reviews have moved to the co-called bridging phase activities. Bridging leads up to the SKA1 construction phase. 
As the primary motivation for the establishments of both AAVS1 and EDA1 was to support the design and prototyping within the LFAA element consortium, both arrays also provide valuable information towards emerging concerns from the LFAA element CDR.
The main focus has been on station calibration, and further prototyping was considered required, looking at the results from AAVS1, combined with advanced insights into the overall SKA-Low architectural design. 


Therefore, within the bridging phase, the prototyping effort has continued into the bridging arrays. These consist of a full station of the next-generation SKALA (SKALA4) antennas (in a station called AAVS2) and a full station of `MWA style' dipoles (in a station called EDA2).
In addition, a 64-element array of SKALA4 antennas is under construction at Lords Bridge Observatory, Cambridge, UK.
These arrays will be used in parallel observation campaigns to validate electromagnetic models and simulations of station calibration and performance. These activities should inform a number of key down-selects and decisions. 

A next phase within Bridging is foreseen, labelled as `Phase 3'. The phase 3 bridging array should consist up to four full field nodes, that is, antenna stations, power and signal distribution systems (PaSD), SPS and external interfaces. 

The antenna stations will hold 256 antennas each, utilising a single, production ready, antenna design. The PaSD deployment and field evaluation is aiming at a fully functional PaSD designed for compliance with the SKA requirement specification. As the SPS deployed for AAVS1 is not representative for the proposed production configuration, the Phase 3 SPS deployment aims at both continuing the assessment, as well as finalising the production specification of the SPS towards procurement. 
The external interface provide an opportunity for testing and verifying the critical interfaces, including telescope manager (TM), the Central Signal Processor (CSP) and the Science Data Processor (SDP).

\section{Conclusion}
\label{sec:conclusions}

We have presented AAVS1, the first full-sized prototype station for SKA-Low. The station was deployed on the MRO site in 2017 and commissioned in 2018 and early 2019.

The station consists of 256 SKALA2 log-periodic antennas and uses RF-over-fibre to send the signals from each antenna over approximately 5\,km of optical fibre to the MRO control building. Signals from the station are processed in groups of 16 antennas by TPMs, which combine analogue and digital components.

The signal processing functionality was commissioned and tested by using the station as an interferometer. This led to offline and real-time testing of the station beam, which is the primary data output produced by an SKA-Low station.

Testing demonstrated that the station could be calibrated reliably in a standalone mode using the Sun between 105 and 308\,MHz, and that the long-term behaviour of the station calibration was stable. Mutual coupling between the antennas was identified as the likely reason why the phases of measured visibilities deviated from ideal behaviour on the timescales of tens of minutes to hours.

The sensitivity of the array was measured at several frequencies by forming difference images of the sky between all-sky snapshot images closely spaced in time. The median sensitivity was 50 per cent under requirement, with the caveat that we only had a very small amount of data to conduct this analysis. The sensitivity results should be considered a solid proof-of-concept of the method, rather than definitive.

The process of procuring, testing, transporting, deploying, and field-testing AAVS1 identified many opportunities for design improvement, as well as recognition that bottom-up component-based costing models can underestimate the true costs and risks of a large-scale deployment.

Building AAVS1 was a large-scale venture undertaken by the AADC Consortium, with goals ranging from validation of the architecture and technology to improving the completeness and maturity of the LFAA cost model. Within this context, AAVS1 should be considered a successful and valuable deployment of the first full-scale SKA-Low prototype station.

\section{Acknowledgement}
This scientific work makes use of the Murchison Radio-astronomy Observatory, operated by CSIRO. We acknowledge the Wajarri Yamatji people as the traditional owners of the Observatory site. AAVS1 lies within the footprint of the Murchison Widefield Array radio telescope and uses power and communications infrastructure from the MWA and MRO site.
Eloy de Lera Acedo acknowledges the support of the Science and Technology Facilities Council (UK).

\bibliographystyle{aa}
\bibliography{refs}
\end{document}